\documentclass[11pt]{article} 

\usepackage{geometry}
\usepackage{microtype}

\usepackage{graphicx}

\usepackage{mathptmx}        
\usepackage{helvet}          
\usepackage{courier}         

\usepackage{amsfonts}
\usepackage{amsmath}
\usepackage{epigraph}
\usepackage{xspace}
\usepackage{mathtools}
\newcommand {\mm}[1] {\ifmmode{#1}\else{\mbox{\(#1\)}}\fi}
\newcommand{\Rspace}        {\mm{{\mathbb R}}}

\newcommand{\Xspace}        {\mm{{\mathbb X}}}

\newcommand{\Mspace}        {\mm{{\mathbb M}}}

\newcommand{\Acal}        {\mm{\mathcal A}}
\newcommand{\Dcal}        {\mm{\mathcal D}}

\newcommand{\Rcal}        {\mm{\mathcal R}}
\newcommand{\Scal}        {\mm{\mathcal S}}

\newcommand{\Hgroup}        {\mm{\sf H}}
\newcommand{\Cgroup}        {\mm{\sf C}}

\newcommand{\bdr}        {\mm{\mathrm{\partial}}}
\newcommand{\rank}        {\mm{\mathrm{rank} }\ }
\newcommand{\myker}        {\mm{\mathrm{ker}}}
\newcommand{\myim}        {\mm{\mathrm{im}}}
\newcommand{\grad}[1]     {{\nabla {#1}}}
\newcommand{\J}        {\mm{{\mathbb J}}}

\newcommand{\para}[1]  {\par\medskip\noindent\textsc{#1. }}

\title{Mathematical Foundations in Visualization}
\author{Ingrid Hotz\footnote{ingrid.hotz@liu.se, Link{\"o}ping University, Sweden}, 
Roxana Bujack\footnote{bujack@lanl.gov, Los Alamos National Laboratory, USA}, 
Christoph Garth\footnote{garth@cs.uni-kl.de, Technische Universit{\"a}t Kaiserslautern, Germany},
Bei Wang\footnote{beiwang@sci.utah.edu, University of Utah, USA}}
\date{}

\begin{document}
\maketitle

\abstract{
Mathematical concepts and tools have shaped the field of visualization in fundamental ways and played a key role in the development of a large variety of visualization techniques.
In this chapter, we sample the visualization literature to provide a taxonomy of the usage of mathematics in visualization, and to identify a fundamental set of mathematics that should be taught to students as part of an introduction to contemporary visualization research.
Within the scope of this chapter, we are unable to provide a full review of all mathematical foundations of visualization; rather, we identify a number of concepts that are useful in visualization, explain their significance, and provide references for further reading.\\
\indent We assume the reader has basic knowledge of linear algebra~\cite{Strang2016}, multivariate calculus~\cite{Steward2019}, statistics, combinatorics, and stochastics~\cite{Georgii2007}.
Other topics not covered in this chapter, such as image analysis~\cite{Sonka2008},
  computer graphics~\cite{Shirley2005},
  signal processing~\cite{Glassner1995},
  computational geometry~\cite{DeBerg1997},
  geometric modeling, mesh generation, computer aided geometric design~\cite{Farin2002,ZhangYJ2016}, and numerics~\cite{Press1992}, can be found in well-established textbooks.
}


\section{Data and basic terminology \index{data}}
\label{sec:data}
\begin{figure}[!ht]
\begin{center}
\includegraphics[width=1\textwidth]{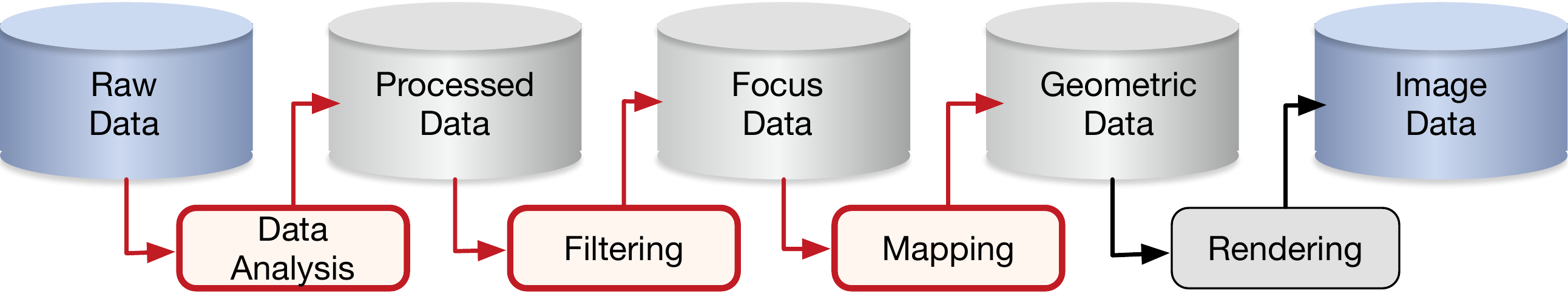} 
\caption{A visualization pipeline. All steps in the pipeline involve the use of mathematical concepts and tools. We cover various aspects of data analysis, filtering, and mapping.}
\label{fig:pipeline}
\end{center}
\end{figure}

\setlength{\epigraphwidth}{0.8\textwidth}
\epigraph{You can have data without information, but you cannot have information without data.}
{\textit{Daniel Keys Moran, programmer and science fiction writer}}
\noindent
Data are at the center of every visualization task and every step of the visualization pipeline, see Fig.~\ref{fig:pipeline}. 

\begin{figure}[!ht]
\begin{center}
 \includegraphics[width=.8\textwidth]{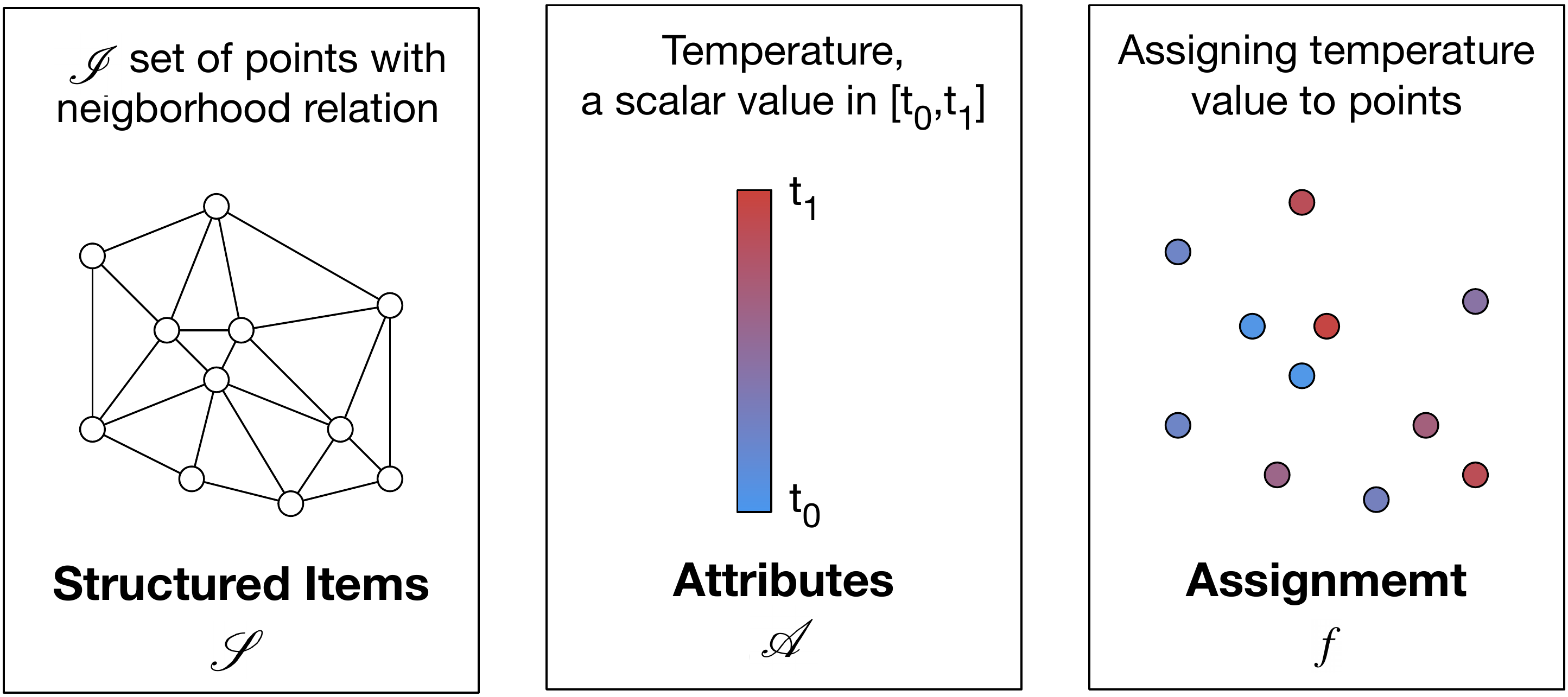} 
\caption{An example of a data set: $\Scal$ consists of a set of points with a neighborhood relation. Attributes in $\Acal$ are elements of the interval $[t_0,t_1]$. $f$ assigns temperature values to the points.}
\label{fig:data_set}
\end{center}
\end{figure}

The input to the visualization pipeline, the raw data, can be any collection of information in any form.  
In this chapter, we define a data set as a triplet $\Dcal = (\Scal,\Acal, f)$ consisting of a set of structured items $\Scal$, a set of attributes $\Acal$, and a function that  assigns attributes to the items.
$\Scal$ consists of a set of items, continuous or discrete, together with a structure (such as a metric for a continuous domain or neighborhood relations for networks), see Fig.~\ref{fig:data_set} for an example.

The tools used for the analysis and visualization of data sets depend on the nature of $\Scal$ and $\Acal$. 
The most important distinctions are continuous vs. discrete structures, and quantitative vs. categorical attributes, see Table~\ref{tab:data}. 
In this section, we emphasize continuous structures and quantitative attributes.
\begin{table}[b]
    \centering
    \begin{tabular}{p{0.55 \textwidth}|p{0.4 \textwidth}}
    Structures  $\Scal$   &  Attributes $\Acal$\\
    \hline
    continuous domains equipped with metrics     &  ordered, ordinal, quantitative \\
    meshes, simplicial complexes &  scalars, vectors, tensors \\
    graphs, networks, trees  & categorical  
    \end{tabular}
    \caption{Examples for possible structures $\Scal$ and attribute spaces $\Acal$.}
    \label{tab:data}
\end{table}
A more detailed classification of data sets concerning types, structures, and organizations can be found in Munzner~\cite{Munzner2014}. 
An introduction to data representations from a scientific visualization perspective can be found in Telea~\cite{Telea2015}. 
\subsection*{The structure $\Scal$}
The structure $\Scal$ can vary from discrete points to continuous domains. 
In general, $\Scal$ consists of a set of items and some relation between the items. We describe two of the most frequently used structures in more detail.

\para{Graphs, networks and trees}
\emph{Graphs}\index{graph} or \emph{networks}\index{network} are structures that are frequently used for non-spatial, relational data representations. 
The terms graph and network are sometimes used interchangeably. 
Mathematically, a graph $G$ is a pair $(V, E )$ consisting of a set of items  $V$, called vertices or nodes, and a set of relationships between these items expressed as a set of edges $E \subseteq V \times V$. 
Edges can be directed or undirected. For directed graphs, $(v,w),$ and $(w,v) \in E$ represent different relations. If the edges are assigned a numeric attribute, the graph is weighted.

A possible representation of a finite graph is an adjacency matrix, which is a square matrix of size $|V| \times |V|$. For a simple graph, the adjacency matrix is a (0,1)-matrix with zeros on its diagonal and ones for each edge. If the graph is undirected, the matrix is symmetric. 
Typically, graphs are displayed using a set of points for the vertices, which are joined by lines for the edges. A general introduction to graphs and networks can be found in~\cite{Jungnickel2012}.

When analyzing graphs, characteristics as cycles, planarity, sparseness, and hierarchical representations are of interest. 

\para{Continuous Domains}
\index{domain - continuous}
 A continuous \emph{domain}\index{domain} $D$ is a subset of $\Rspace^n$ equipped with a \emph{metric}.
 A metric supports measurements and determines distances in the domain.
 A common metric is the Euclidean distance. 
 Other metrics include Manhattan distances and polar distances. 
 More generally, when the domain is a \emph{parameterized manifold}, the choice of a metric has an impact on many calculations such as derivatives, see Section~\ref{sec:differential}. 
 
A continuous domain can be represented by a finite set of discrete samples associated with an interpolation scheme. 
In this case, $\Scal$ consists of a  set of points $\{p_i\in D \mid i = 1,\dots, k\}$, equipped with a neighborhood structure; e.g.,  the points are organized as a \emph{regular grid}\index{regular grid} (associated with piecewise multilinear interpolation) or a simplicial complex (corresponding to piecewise linear interpolation). 
\subsection*{The attribute space $\Acal$}
\label{suvsec:attributeSpace}
An attribute is a specific property assigned to data items that arise from measurement, observation or computation. 
Attributes can be continuous and quantitative, e.g.,~temperature; discrete and ordered, e.g.~the number of people in a class; as well as categorical, e.g.,~various types of tree species. The set of possible attributes span the \emph{attribute space}\index{attribute space}. 

The most common \emph{continuous quantitative attributes}\index{attributes - continuous quantitative} can be subsumed under the term \emph{tensor}. 
A \emph{tensor of order $r$}\index{tensor}\index{tensor order} is defined as a multi-linear mapping acting on $r$ copies of a $n$-dim vector space $V$ over $\Rspace$  into the space of real numbers, 
\begin{equation}
            \mathbf{T}: \underbrace{V \times \ldots \times V}_{r} \rightarrow \Rspace.
    \label{eq:tensor}
\end{equation}
Sometimes \emph{rank}, \emph{degree} and \emph{order} are used interchangeably. 
A tensor of order $0$ corresponds to a scalar $\alpha \in \Rspace$ and a  tensor of order $1$ is a vector  $\mathbf{v}\in V$. 
\begin{eqnarray*}
            \alpha:   \Rspace\rightarrow \Rspace, & \alpha(x)=\alpha x & \mbox{$0$th-order tensor or \emph{scalar}} \mbox{, e.g.,~temperature;} \\
            \mathbf{w}: V  \rightarrow \Rspace, & \mathbf{w}(\mathbf{v})=\mathbf{w}\cdot\mathbf{v} & \mbox{$1^{st}$-order tensor or \emph{vector}} \mbox{, e.g.,~velocity;}\\
            \mathbf{T}: V\times V  \rightarrow \Rspace, & \mathbf{T}(\mathbf{v,v'})=\mathbf{v}\cdot\mathbf{T}\cdot\mathbf{v'} & \mbox{$2^{nd}$-order \emph{tensor}, e.g., strain tensor.}
\end{eqnarray*}
\emph{Tensors of higher order}\index{tensor of higher order}, especially $3^{rd}$- and $4^{th}$-order tensors, can also be found in a few visualization applications.
In the visualization literature, the term tensor often refers to $2^{nd}$-order tensors.
With respect to a specific basis  $\{\mathbf{e}_1, \ldots, \mathbf{e}_n\}$
of the vector space $V$,  a tensor is fully specified by its action on the basis elements resulting in the typical component representations. For a vector, this is 
$w = (w_1,\dots,w_n)^T$
and for a $2^{nd}$-order tensor, this is a matrix 
\[\mathbf{T}=
\left(
\begin{array}{ccc}
t_{1,1} & \cdots & t_{1,n} \\  \vdots & \ddots &\vdots    \\t_{n,1} & \cdots & t_{n,n}
\end{array}
\right).\]
For a basic introduction to the use of tensors in visualization, we refer to the state-of-the-art report by Kratz et al.~\cite{Kratz2013}.

\para{Enriched Attribute Space $\Acal^*$}\index{enriched attribute space}
In-depth data analysis often requires some modifications of the attribute space. The most common examples are \emph{filtering}\index{filtering}, e.g. removing noise, or \emph{enrichment}\index{enrichment} of the original attributes by derived quantities, e.g. the field gradient or local histograms. 
Other modifications are changes of the representation or parameterization  of the attribute space to emphasize data symmetries useful for feature or pattern definitions; see also Section~\ref{sec:features}. Examples include scaling, rotation in attribute space, and expressing a $2^{nd}$-order tensor by its eigenvalues and eigenvectors.
\subsection*{Fields as example data sets}
Field data are very common in scientific applications where they  express physical quantities defined over continuous domains, for instance, temperatures in a room, or wind velocities in the atmosphere.
Such data are often the results of numerical simulations or measurements from experiments.
A \emph{field}\index{field} is defined as a mapping from a continuous \emph{domain} $D\subseteq \Rspace^n$ into an \emph{attribute space} $\Acal \subseteq \Rspace^m$ (similar notions include \emph{range}\index{range} and \emph{co-domain}\index{co-domain}), given as 
\begin{equation}
  f:\Rspace^n \supseteq D \rightarrow \Acal\subseteq \Rspace^m.
  \label{eq:field}
\end{equation}
Typically, the domain can be considered in a spatiotemporal context, for example, $D= D_s\times I_t \subseteq  \Rspace^4$, where $D_s \subseteq \Rspace^3$ is the spatial domain and $I_t \subseteq \Rspace$ is a time interval.
Depending on the attribute space, we distinguish a 
\emph{scalar field}\index{scalar field} $S: D \to \Acal \subseteq \Rspace$,
a \emph{vector field}\index{vector field} $V: D \to \Acal \subseteq \Rspace^2$, 
a \emph{tensor fields}\index{tensor field} $T$, and 
more generally, a combination of such fields, resulting in a \emph{multifield}\index{multi-field} with an attribute space spanned by the individual fields.

\para{Ensembles of Fields}
Fields are often associated with a set of parameters, which typically play a different role than the domain dimensions. 
Parameters are often used to create collections of data sets, referred to as \emph{ensembles}\index{ensembles}~\cite{WangJ2018}.
\begin{equation}
  \{f_1, \cdots, f_k\}: D\times\{P_1,\dots,P_k\}\rightarrow \Acal\subset\Rspace^m,\newline
  \label{eq:ensemble}
\end{equation}
where each $P_i$ (for $i=1\dots k$) is a parameter tuple. 
An example of an ensemble is the data set generated from a computer simulation with different initial conditions (described by different parameters). 
Each $f_i: D \times P_i \to \Acal$ is an \emph{ensemble member} or a \emph{realization}. 
Ensemble members often have internal correlations or follow certain distributions, making them especially hard to analyze. 
Ensemble data arise in many applications and is an important theme in visualization research~\cite{Hansen2014}.

\section{Differential structures}
\label{sec:differential}
\begin{figure}[t]
\begin{center}
\begin{tabular}{ccc}
 \includegraphics[height=.34\textwidth]{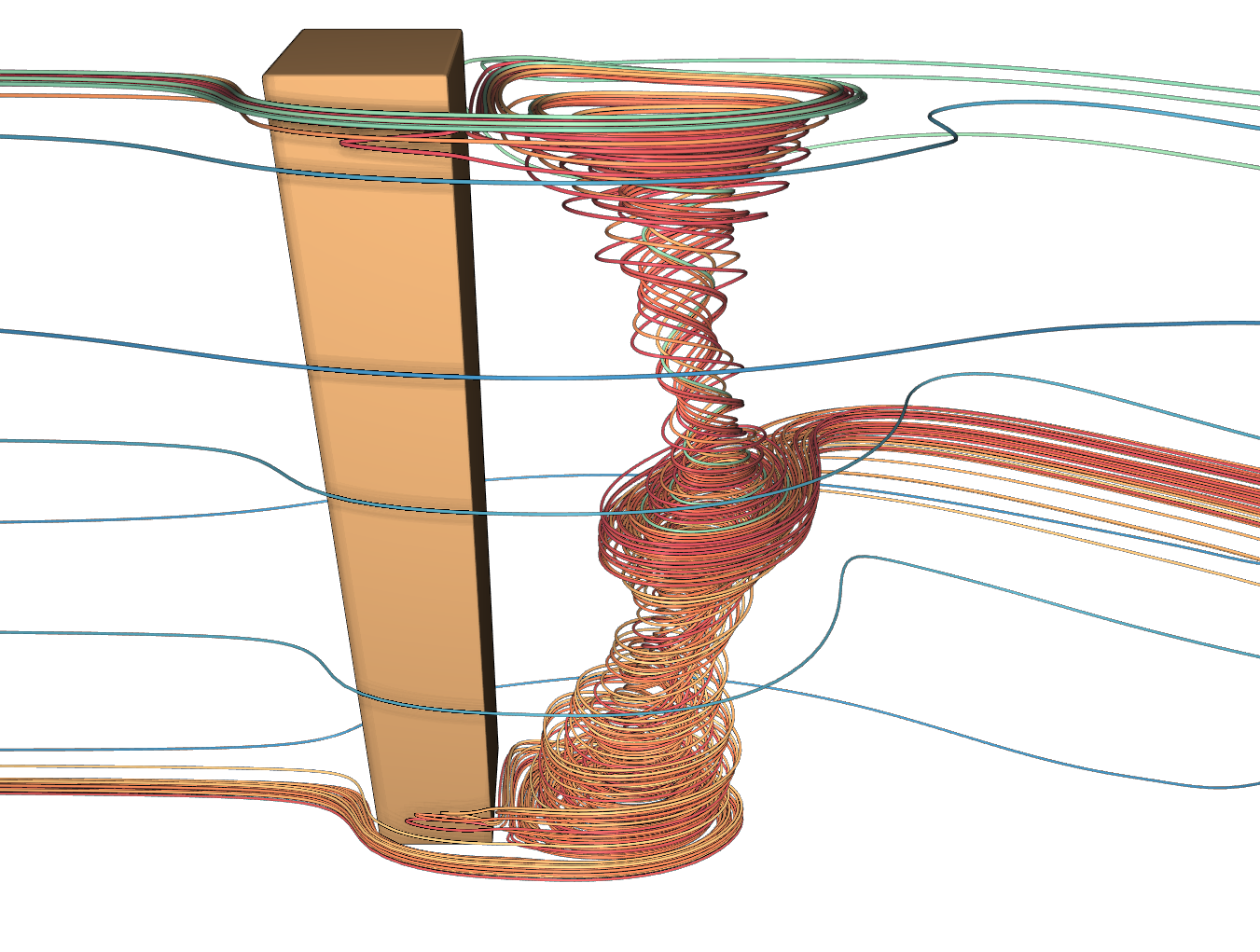} &&
 \includegraphics[height=.35\textwidth]{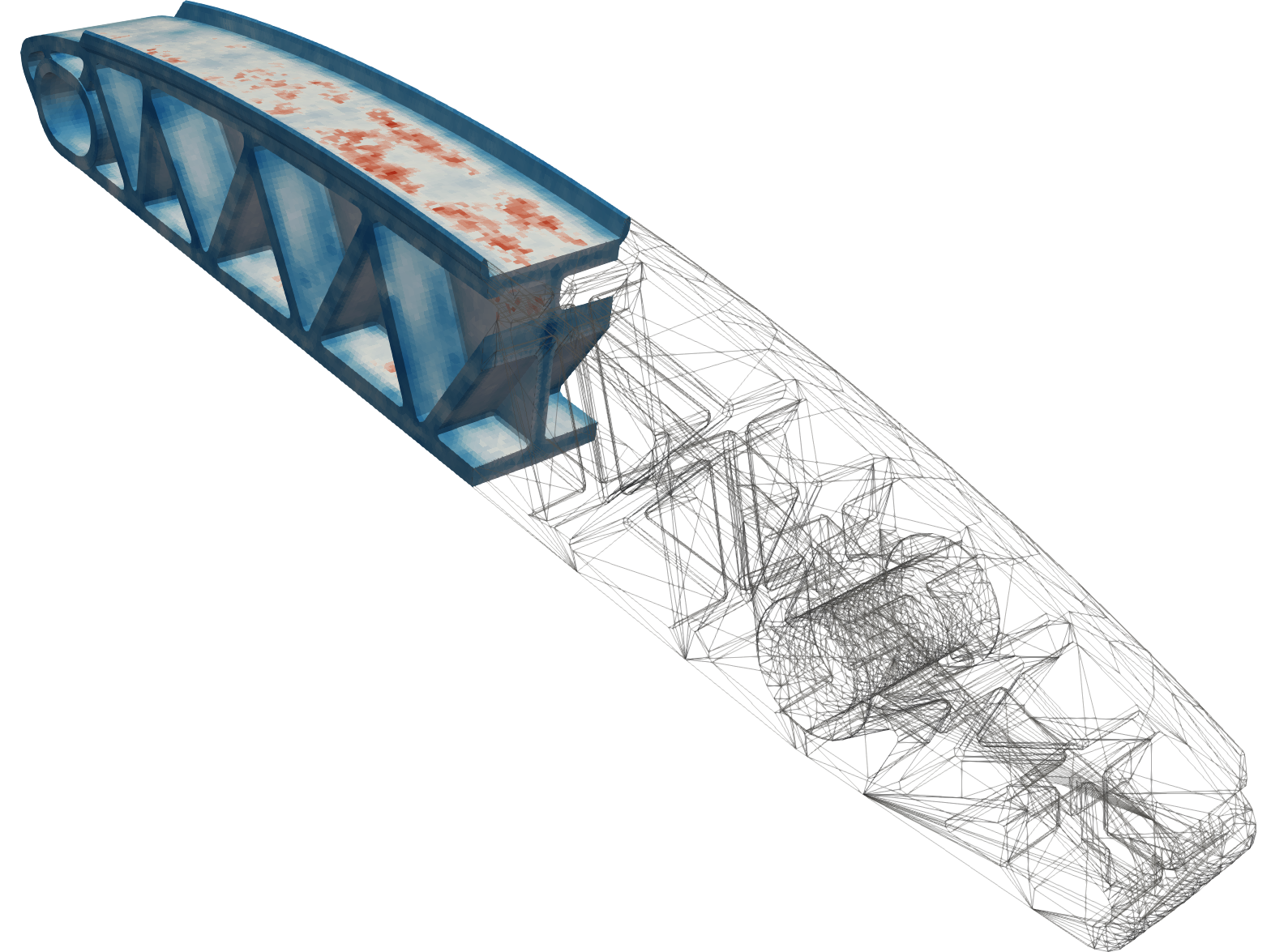}\\
 (a) && (b)
 \end{tabular}
\caption{An interplay between discrete data and continuous concepts.  (a) Numerically computed streamlines of the flow behind a cylinder approximate the solutions of an ordinary differential equation. (b) A discrete mesh approximates the shape of a mechanical part, where a continuous color map highlights the extremal values of the load of the material.}
\label{fig:discrete-continuous}
\end{center}
\end{figure}

\setlength{\epigraphwidth}{0.8\textwidth}
\epigraph{Science is a differential equation. Religion is a boundary condition.}
{\textit{Alan Turing, mathematician and computer scientist}} 
\noindent
Whereas real data and computations are mostly based on discrete domains and attributes, many of the concepts for their analysis are founded on continuous settings.
The machinery of differential arithmetics and differential structures  provide powerful analysis tools. 
\emph{Differential operators}~\cite{coffin1911vector, snider1987introduction} play a crucial part in visualization. They allow the definition and categorization of many features, including \emph{extrema}, \emph{ridges}, \emph{valleys}, \emph{saddles} and \emph{vortices}. \emph{Differential equations}, for example, are the basis for the definition of streamlines, a fundamental method in flow visualization, see Fig.~\ref{fig:discrete-continuous}(a). 

Finally, \emph{differential geometry} provides mathematical tools to characterize curves and surfaces and plays an important role in visualization, see Fig.~\ref{fig:discrete-continuous}(b). 
In this chapter, we summarize the most fundamental concepts of discrete structures that are frequently encountered in visualization research.  

\subsection*{Differential operators}
\para{Differential operators in Euclidean spaces}\emph{Differential operators}~\cite{coffin1911vector, snider1987introduction} 
 map functions (e.g. fields) to their derivatives and thus allow us to study the rates at which continuous attributes change. They can be applied to scalar, vector, and tensor fields.
They give rise to definitions of features, such as \emph{extrema}, \emph{ridges}, \emph{valleys}, \emph{saddles}, and \emph{normals} of isosurfaces. 
We describe differential operators for scalar fields $f:\Rspace^n \to \Rspace$ and vector fields $v:\Rspace^n \to \Rspace^n$.
The explicit expression of the operators depends on the inherent metric of the space; here, we assume the Euclidean metric.
We often use the operator \begin{equation} \begin{aligned}
\nabla=\begin{pmatrix}
\frac{\partial}{\partial x_1}\\...\\\frac{\partial}{\partial x_n}
\end{pmatrix}
\end{aligned} \end{equation}
to simplify the notations.
The \emph{gradient} of a scalar field 
\begin{equation} \begin{aligned}
\nabla f = \begin{pmatrix}
\frac{\partial f}{\partial x_1}\\...\\\frac{\partial f}{\partial x_n},
\end{pmatrix}
\end{aligned} \end{equation}
 is a vector that indicates the direction of the steepest ascent.
 Locations where gradient vanishes ($\nabla f =0$) are associated with \emph{critical points}\index{critical point scalar field} of the scalar field, such as maxima, minima, and saddles, see also Section~\ref{sec:topology}. 
\emph{Hessian matrices}\index{Hessian matrix} consisting of $2^{nd}$-order partial derivatives are used to classify the critical points, 
\begin{equation} \begin{aligned}
H =  \begin{pmatrix}
\frac{\partial^2 f}{\partial x_1^2} &...& \frac{\partial f}{\partial x_1\partial x_n}\\ 
...&&... \\ 
\frac{\partial^2 f}{\partial x_n\partial x_1}&...&\frac{\partial^2 f}{\partial x_n^2}
\end{pmatrix}.
\end{aligned} \end{equation}
The eigenvalues of the Hessian $H$ can be interpreted as the principal curvatures, and the eigenvectors as principal directions;
therefore $H$ is often used to define ridges and valley lines in scalar fields. 
For example, a \emph{topographic ridge} is defined as the set of points where the slope is minimal on the scalar field restricted to a contour line. 
This means that one eigenvector of $H$ is aligned with the elevation gradient~\cite{Peikert1999}.

\begin{figure}[!ht]
\begin{center}
\includegraphics[width=.9\textwidth]{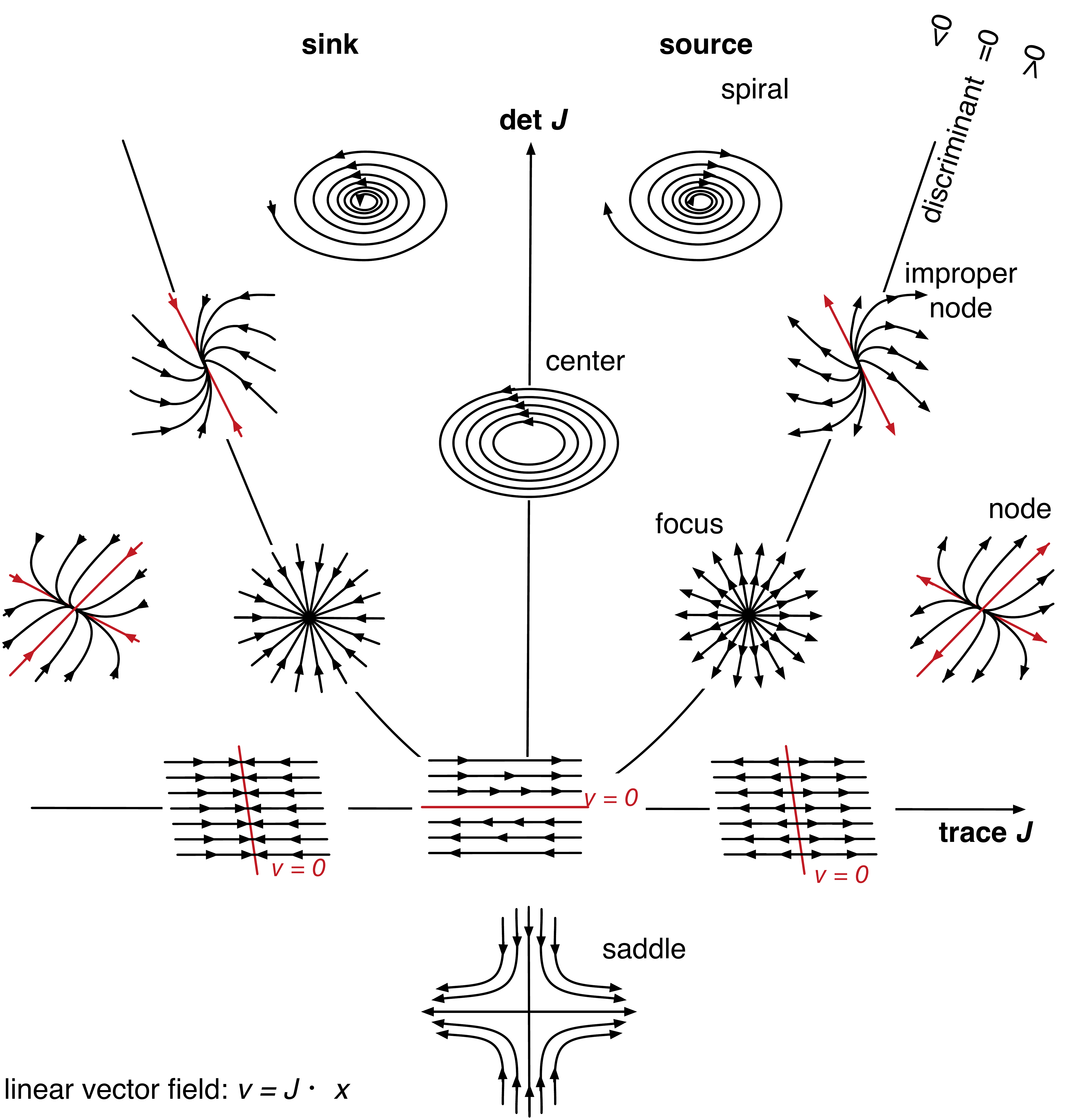}
\caption{The Jacobian $J$ can be used to classify the local behavior of a vector field in the vicinity of a critical point. Locally, the field can be approximated up to $1^{st}$ order via the Taylor expansion as $v( x)=v_0+J\cdot x$. If $v_0=0$, the point $ x$ is critical. The critical point can be classified based on the determinant and the trace of the Jacobian. The sign of the discriminant $\Delta =\text{tr}^2(J)- 4\det(J)$ separates the area of real and complex eigenvalues of the Jacobian. Complex eigenvalues are associated with swirling motions. }
\label{fig:critical_classification}
\end{center}
\end{figure}

The \emph{Jacobian} $J\in \Rspace^{n\times n}$ is  a matrix that generalizes the concept of a gradient for a vector field $v$,
\begin{equation} \begin{aligned}
J =\nabla v = \begin{pmatrix}
\frac{\partial v_1}{\partial x_1}&...&\frac{\partial v_n}{\partial x_1}\\
...&&...\\
\frac{\partial v_1}{\partial  x_n}&...&\frac{\partial v_n}{\partial x_n}
\end{pmatrix}.
\end{aligned} \end{equation}
The eigenvalues of the Jacobian can be used to categorize the types of $1^{st}$-order \emph{critical points}\index{critical point vector field} in vector fields, i.e.,~positive for sources, negative for sinks, differently signed for saddles, and complex for center points, see   Fig.~\ref{fig:critical_classification}.

Other important differential operators are 
the \emph{Laplace operator} 
$\Delta f = \nabla^2f =
\frac{\partial^2 f}{\partial x^2} + \frac{\partial^2 f}{\partial y^2} + \frac{\partial^2 f}{\partial z^2}$,
the \emph{divergence}
$\operatorname{div} v =\nabla\cdot v $,
and the \emph{curl}
$\operatorname{curl} v =\nabla\times v$ 
of a vector field. In an infinitessimal neighborhoord, the divergence is a measure of how much the flow converges toward or repels from a point, and the curl indicates of how much the flow swirls or rotates. 

\para{Differential operators for field approximations} 
Differential operators also play an important role in the approximation of fields as they represent the components in the \emph{Taylor expansion}. A scalar field in the vicinity of a point $P\in\Rspace^n$ can be approximated as  $f(P+x)=f(P)+\nabla f (P)\cdot x + \frac{1}{2}x^T H(P) x +O(\|x\|^3)$. For vector fields, the linear approximation is given as
$v(P+x)=v(P)+J(P)\cdot x +O(\|x\|^2)$.

\para{Differential operators in non-euclidean spaces}
For non-Euclidean spaces, differential operators are more complex. 
Consider, for example, spherical coordinates: 
the divergence of a vector $(v_r, v_\theta, v_\varphi)$ (where $r$ is the radius, $\theta$ is the polar angle, and $\varphi$ is the azimuthal angle) is then given as
\begin{equation}
\operatorname{div} v  = 
\frac{1}{r^2}\frac{\partial \left( r^2 v_r \right) }{\partial r}
+ \frac{1}{r\sin\theta}\frac{\partial}{\partial \theta} \left(  v_\theta\sin\theta \right)
+ \frac{1}{r\sin\theta}\frac{\partial v_\varphi}{\partial \varphi}.
\end{equation}
The differential operators for cylinder and spherical coordinates can be found in most textbooks.

\subsection*{Differential equations}
A \emph{differential equation}~\cite{amann2011ordinary,renardy2006introduction} is a mathematical equation that relates a function with its derivatives. 
Differential equations are categorized into ordinary differential equations (containing one independent variable), and partial differential equations (involving two or more independent variables). 

One of the most common examples of an ordinary differential equation in visualization is given through the relation of a vector field and its trajectories (that are everywhere tangential to the field), see  Fig.~\ref{fig:discrete-continuous}(a).
A flow can be represented either as a time dependent vector field
$\Rspace^d\times \Rspace\to\Rspace^d,$ $( x,t)\mapsto  v(x,t)$
or through its \emph{flow map},
 \begin{equation}\begin{aligned}
\Rspace\times \Rspace\times \Rspace^d\to\Rspace^d,&& t\times t_0\times x_0\mapsto F_{t_0}^t(x_0),
\end{aligned}
\end{equation}
\[\begin{aligned}
\mbox {with } F_{t_0}^{t_0}(x_0)= x_0, & \mbox{  and }& 
  F_{t_1}^{t_2}(F_{t_0}^{t_1}(x_0))=&F_{t_0}^{t_2}(x_0).
\end{aligned}
\]
The flow map describes how a flow parcel at $(x_0,t_0)$ moves to $ F_{t_0}^{t_{1}}(x_0)$ in the time interval $[t_0, t_1]$. 
The two representations of the vector field are related through the initial value problem~\cite{coddington2012introduction},  
 \begin{equation}\begin{aligned}\label{FMDiff}
\dot F_{t_0}^{t}(x_0) = v(x(t), t)&,& F_{t_0}^{t_0}(x_0)=x_0, 
\end{aligned}\end{equation}
where $\dot F$ refers to the temporal derivative of $F$, and inversely through integration, 
 \begin{equation}\begin{aligned}\label{VFIntergral}
x_0+\int_{t_0}^{t}v(x(t),t) d t = F_{t_0}^{t}(x_0).
\end{aligned}\end{equation}
Partial differential equations are more complex than ordinary differential equations, and, depending on the initial and boundary conditions~\cite{folland1995introduction}, may not have a unique solution or a solution at all. As a popular example, we can look at the heat equation, 
 \begin{equation}\begin{aligned}
\frac{du(x,t)}{dt} -\alpha \nabla^2 u(x,t) =0, 
\end{aligned}\end{equation}
Where $\alpha\in\Rspace$ is called the \emph{thermal diffusivity}. 
The solution to the above heat equation is a Gaussian.
It describes the physical problem of heat transfer or diffusion and is used in various visualization applications, for instance, in diffusion-based smoothing, or to define a continuous scale space. 

Even if solutions of differential equations exist, for visualization applications, it is rarely possible to derive them analytically, but only numerically~\cite{butcher2016numerical, morton2005numerical}, due to the reliance on empirical data for coefficients, initial conditions, and boundary conditions. The most popular solvers for ordinary differential equations are the Euler and Runge-Kutta methods. For partial differential equations, the families of finite element methods (FEM), finite volume schemes, and finite differences methods are frequently used, depending on the choice of discretization.

\subsection*{Differential geometry}
We review elements from differential geometry~\cite{kuehnel2015differential} that are most relevant to visualization, including parametrized curves and surfaces, lengths, areas, and curvature. 
 Some of these concepts can be generalized from three-dimensional to higher-dimensional spaces dealing with general manifolds, which are topics in Riemannian Geometry~\cite{Berger2003}.
 
\para{Parametric curves}
In differential geometry, curves are defined in a parametrized form, and their geometric properties, including arc length, curvature, and  torsion, are expressed using integrals and derivatives. 
 A \emph{parametric curve}\index{parametric curve} 
 \begin{equation}
     \gamma:[a,b]\subset\Rspace\to\Rspace^d
 \end{equation}
 is a vector-valued function defined over a non-empty interval.
Curves can be distinguished depending on how often they are differentiable. In the continuous case, we will assume the curve to be sufficiently smooth.

The fundamental theorem of differential geometry of curves guarantees that up to transformations of the Euclidean space (rotations, reflections, and translations), a three-dimensional curve can be uniquely defined by its \emph{velocity}, \emph{curvature}, and \emph{torsion}. 
These three concepts describe changes of the \emph{Frenet-Serret frame}\index{Frenet frame}, which is a local coordinate system that moves with the curve. 
A Frenet-Serret frame is spanned by the unit \emph{tangent} vector $T(t)$, \emph{normal} vector $N(t)$, and \emph{binormal} vector  $B(t)$, which are defined via derivatives of the curve $\gamma(t)$ with respect to the parameter $t\in[a,b]$, 
    \begin{align*}
T(t) & =\frac{\gamma'(t)}{\|\gamma'(t)\|},  \\
 N(t) & =\frac{\gamma''(t)-\left(\gamma''(t)\cdot T(t)\right) T(t)}{\|\gamma''(t)-\left(\gamma''(t)\cdot T(t)\right)\|},  \\
B(t) & = T(t)\times N(t) .
    \end{align*}
Consequently, commonly used curve descriptors include   \emph{velocity} $v(t)=\|\gamma'(t)\|$,  \emph{curvature} $\kappa(t)={\|T'(t)\cdot N(t)\|}/{\|T(t)\|}$, and \emph{torsion} $\tau(t)={\|N'(t)\cdot B(t)\|}/{\|T(t)\|}$.
\index{speed}\index{curvature}\index{torsion}
Other useful measures are the \emph{arclength}\index{arclength}
$l(t)=\int_a^t\|\gamma'(s)\|ds$ and the \emph{acceleration}\index{acceleration} 
$a(t)=\gamma''(t)$.

\begin{figure}[t]
\centering
\includegraphics[width=1\textwidth]{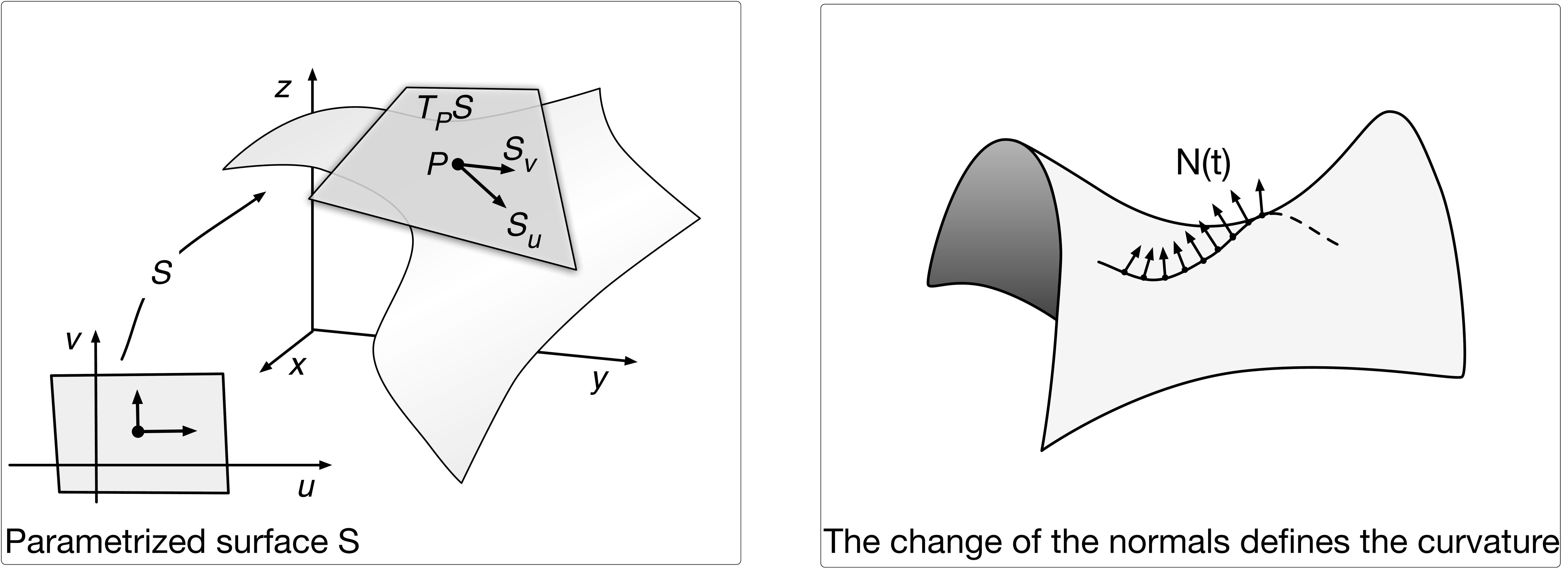}
\caption{Left: parametrized surface. Right: The changes of the normals in a certain direction define the normal curvature of the surface.}
\label{fig:coordinates}
\end{figure}

\para{Parametric Surfaces}
Similar to curves, surfaces can be parametrized, see Fig.~\ref{fig:coordinates}.
 A parametric surface 
 \begin{equation}
     S:\Omega\subset\Rspace^2\to\Rspace^n
\end{equation}
     is a vector-valued function of a non-empty area.
We assume the surface to be sufficiently smooth.

The \emph{tangent plane} \index{tangent plane} of a surface at a point $S(p)\in \mathbb{R}^n$ with $p\in\Rspace^2$ is the union of all tangent vectors of all curves through $S(p)$. 
The plane is spanned by the two partial derivatives $S_u(p)=\partial S/\partial u$ and $S_v(p)=\partial S/\partial v$. The \emph{surface normal}\index{surface normal}, perpendicular to the tangent plane, is given by the cross product of the partial derivatives, 
\begin{equation} 
\begin{aligned} 
\label{surfaceNormal}
 N(p)=\frac{S_u(p)\times S_v(p)}{\|S_u(p)\times S_v(p)\|}.
\end{aligned} 
\end{equation}

\para{Measurements on surfaces} The calculation of the length of a curve on a surface or the surface area can be easily formulated using the \emph{first fundamental form} $\mathrm{I}(p):\Rspace^2\to \Rspace^{2\times 2}$. 
$\mathrm{I}(p)$ defines a natural local metric induced by the Euclidean metric in $\Rspace^n$.  
For notational simplicity, we omit the dependence of the location $p\in\Rspace^2$. Its components $g_{uv}$ are defined as the scalar product of the tangent vectors $S_{u}\cdot S_{v}$. In matrix form, the first fundamental form is given as, 
\begin{equation}
\mathrm{I}=
\begin{pmatrix}g_{{uu}}&g_{{uv}}\\g_{{vu}}&g_{{vv}}\end{pmatrix}
=\begin{pmatrix}E&F\\F&G\end{pmatrix}.
\end{equation}
Using the first fundamental form, a line element $ds$ on the surface is expressed as  $ds^{2}=Edu^{2}+2Fdudv+Gdv^{2}$
and an area element as $dA=\|S_{u}\times S_{v}\| du\,dv={\sqrt  {EG-F^{2}}}\,du\,dv$. 
The arclength of a curve on the surface results from integrating the line element $l=\int_a^bds $, and the area of a surface patch results from integrating the area element. 
 
\para{Surface curvature} 
Many different curvature measures are available.
Loosely speaking, curvature is a concept that measures the amount by which a surface deviates from a plane or the variation of the surface normal. Central to the concept of curvature is the \emph{Gauss map}\index{Gauss map},
which maps the surface normals to the unit sphere $N:S\rightarrow S^2$.  The differential of the Gauss map in a certain direction is a measurement of curvature in that direction. 
Mathematically, the curvature is summarized in the \emph{second fundamental form}\index{second fundamental form}, denoted as $\mathrm{I\!I}$. In matrix form, it is given as, 
\begin{equation}
\mathrm{I\!I}=
\begin{pmatrix}S_{{uu}}\cdot N & S_{{uv}}\cdot N \\ S_{{vu}}\cdot N & S_{{vv}}\cdot N\end{pmatrix}
=\begin{pmatrix}e&f\\f&g\end{pmatrix},
\end{equation}
where $S_{uv},S_{uu},S_{vv}$ are the respective second derivatives of the the surface parametrization. The \emph{shape operator}\index{shape operator} expresses the curvature in \emph{local coordinates}, 
\begin{equation}
    S=\frac{1}{EG-F^{2}}{\begin{pmatrix}eG-fF&fG-gF\\fE-eF&gE-fF\end{pmatrix}}.
\end{equation}
Its eigenvalues ($k_1$ and $k_2$) are called the \emph{principal curvatures}\index{principal curvature} at a given point; and its eigenvectors are called the  \emph{principal directions}\index{principal direction}.
The \emph{Gaussian curvature}\index{Gaussian curvature} $K$ is equal to the product of the principal curvatures. It can also be calculated as the ratio of the determinants of the second and first fundamental forms. The \emph{mean curvature}\index{mean curvature} $H$ is defined as the average of the principal curvatures:
\[K=k_1\cdot k_2=\frac{eg-f^2}{EG-F^{2}}, \hspace{1cm}
  H=\frac{1}{2}(k_1 + k_2)=\frac{1}{2} \frac{eG+gE-2fF}{EG-F^{2}}.\]
Points on the surface can be categorized as \emph{elliptic} ($K>0$), \emph{parabolic} ($K=0,H\neq0$), \emph{hyperbolic} ($K<0$), and \emph{flat} ($K=H=0$ using the Gaussian and mean curvatures.

The curvature $\kappa$ of a surface curve $\gamma$ can be decomposed into its \emph{normal curvature}\index{normal curvature} $k_n$ normal to the surface and its \emph{geodesic curvature}\index{geodesic curvature} $k_g$, which measures the deviation of a curve from being a geodesic $\kappa^2  =k^2_n+k^2_g$.
The extrema of the normal curvature over all curves through a point correspond to the principal curvatures $k_1$ and $k_2$ of the surface. 
A curve where the geodesic curvature is equal to zero is called a \emph{geodesic}\index{geodesic}, which is a generalization of a straight line on arbitrary surfaces, as the straightest and locally shortest curve.

\subsection*{Manifolds}
Roughly speaking, an \emph{n-manifold}\index{n-manifold} $M$ embedded in $\Rspace^m$ is a space that is locally similar to Euclidean space $\Rspace^n$.
Formally, each point $p$ of the manifold $M$ has an open neighborhood $U_p\subset M$ that is homeomorphic to an open subset $V$ of the Euclidean space described by a chart or local frame
$\varphi:U_{p}\subset M \rightarrow V_{p}\subset\Rspace^n$.
The entire manifold can be described by a collection of compatible charts, which together form an atlas. 

A well-known example is a sphere, which is a 2-manifold embedded in $\Rspace^3$, defined by the condition $x^2+y^2+z^2=R$ ($R$ being the radius). There are many ways to define charts on the sphere. It is also possible to cover the whole sphere excluding one point with a chart, which requires at least two charts to complete the atlas.  Covering a sphere with one chart, however, is not possible. 

\begin{figure}[t]
\centering
 \includegraphics[width=1\textwidth]{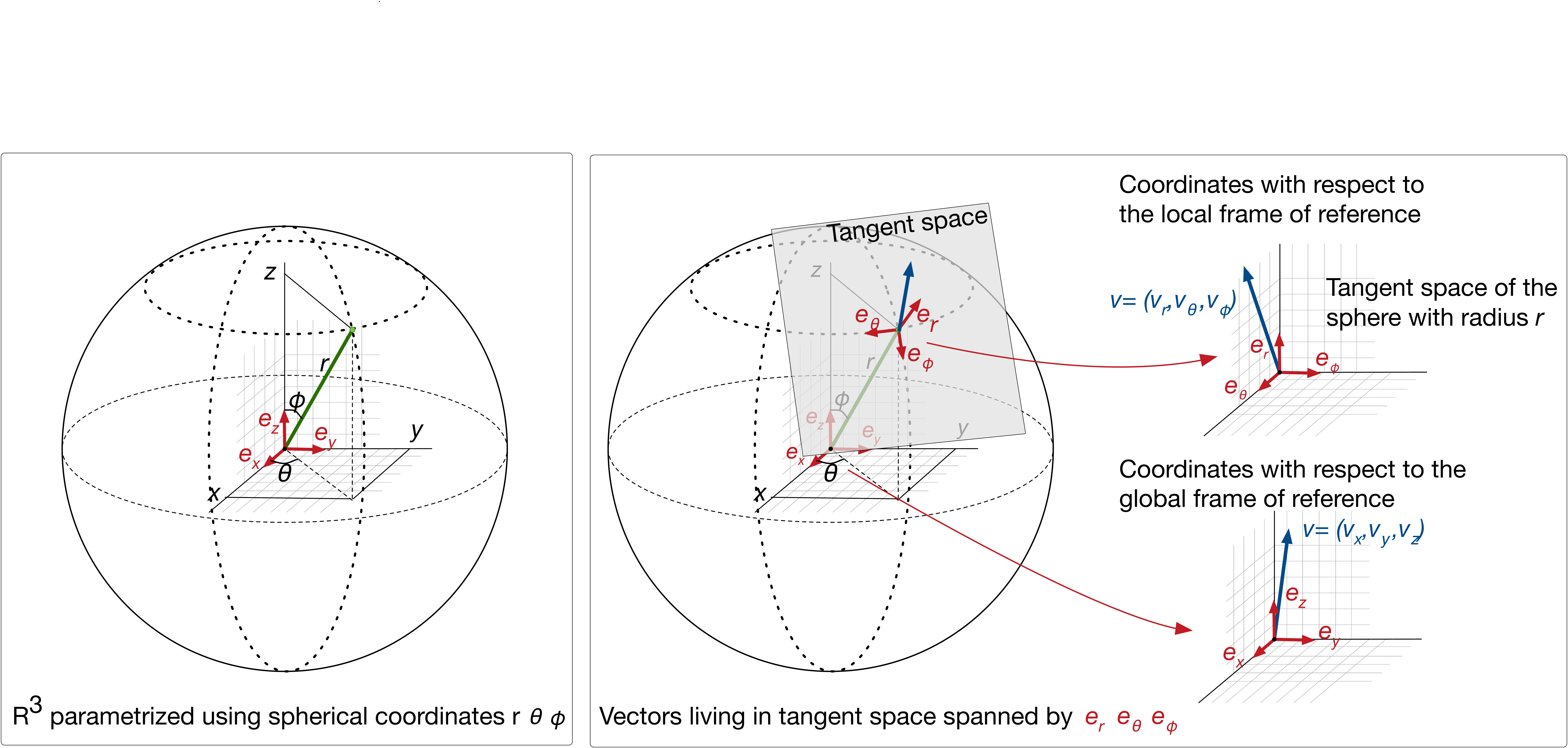}
\caption{Vector field defined on a sphere given in spherical coordinates. Left: a parametrization of the sphere with spherical coordinates, Right: the vector field can be expressed in a local reference frame, which depends on the spherical coordinates.}
\label{fig:manifolds}
\end{figure}

Similar to surfaces, one can define a tangent space $T_pM$ attached to every point in $M$. $T_pM$ has the same dimension as the manifold. The tangent space defines a local basis on the manifold and plays an important role since many fields (e.g. vector fields)  live in the tangent space of the domain, see Fig.~\ref{fig:manifolds}.

\section{Sampled Data and Discrete Methods}
\label{sec:discrete}
\setlength{\epigraphwidth}{0.8\textwidth}
\epigraph{The world is continuous, but the mind is discrete.}
{\textit{David Bryant Mumford, mathematician}} 
\noindent
Fields are defined over continuous domains in theory; however, they are described at discretely sampled locations in practice. 
Typical analysis and visualization methods rely on a \emph{reconstruction}\index{reconstruction} of the continuous fields.  
Two different approaches are commonly used to deal with this issue. 
First, the discrete data is  interpolated to fill the entire domain.
Second, the analysis techniques are transferred to the discrete setting.

\subsubsection*{Data representation}
\begin{figure}[b]
\centering
\includegraphics[width=0.9\textwidth]{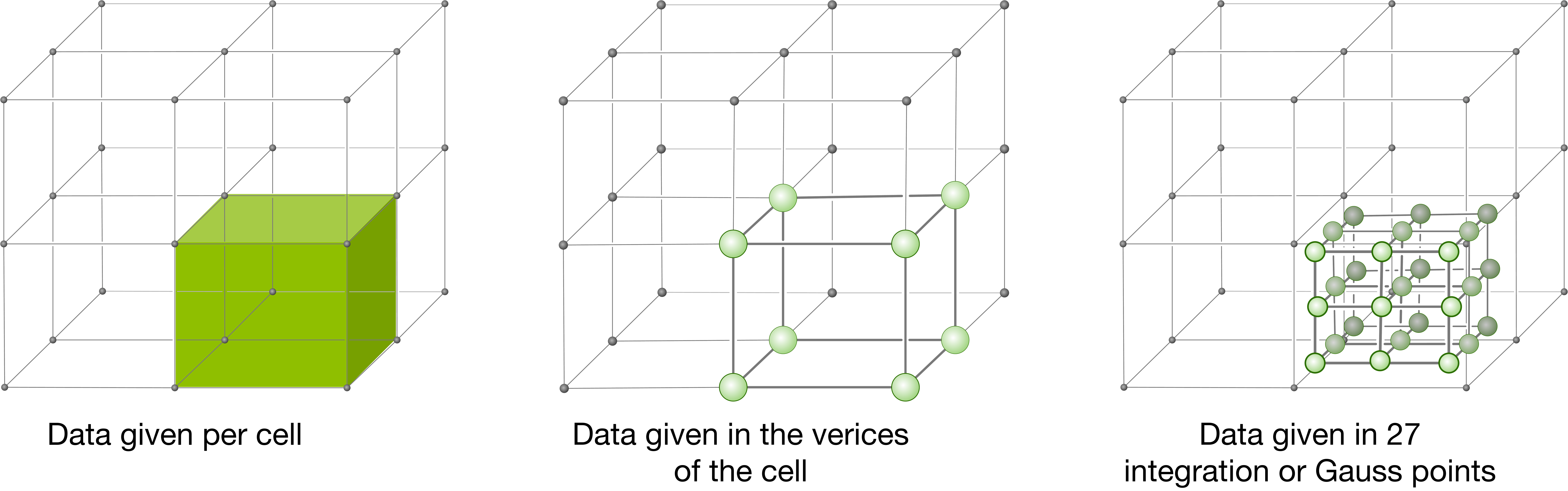}
\caption{Data can be assigned to a regular cubic grid in many different ways.}
\label{fig:data_structure}
\end{figure}

\index{data representation}
Sampled data come in many different forms and representations depending on their origin. For measurement data, one often deals with \emph{unstructured point clouds}\index{point cloud} resulting from practical constraints, e.g.,~possible placements for sensors. 
Data coming from simulations are mostly based on grid structures, ranging from \emph{uniform grids} to \emph{unstructured} and \emph{hybrid grids}\index{uniform grid}\index{unstructured grid}. 
Therefore, the attributes are assigned to either the grid vertices, the grid cells, or distinguished points inside the cells, e.g.,~\emph{Gauss} or \emph{integration points}\index{Gauss point} coming from finite element simulations, see Fig.~\ref{fig:data_structure}.
An overview of common data representations can be found in~\cite{Telea2015}.

A grid is built from a set of vertices $V$ and neighborhood relations,  defining edges, faces and cells.
The neighborhood relations can be given explicitly for unstructured grids or implicitly encoded in an index structure. An example is a quad mesh where the vertices are identified by three indices $V=\{v_{ijk} \mid 1 \leq i,j,l \leq n\}$ and edges $E=\{(v_{ijk},v_{i+1jk}),(v_{ijk},v_{ij+1k}),(v_{ijk},v_{ijk+1}) \mid \forall v_{ijk} \in V\}$.
The most common $2$-dimensional cells are triangles and rectangles; $3$-dimensional cells include quads, tetrahedra, and prisms.

\subsubsection*{Simplicial complexes}
\emph{Simplicial complexes}\index{simplicial complex} are data structures that are particularly useful for combinatorial algorithms (see Section~\ref{sec:topology}). 
They can be considered as a formal generalization of triangulations to higher dimensions. 
A $k$-simplex is defined as the convex hull of $k+1$ affinely independent points $p_i\in\Rspace^k$; the convex hull of any nonempty subset of the $k+1$ points is a \emph{face} of the simplex.
$0$-, $1$-, $2$- and $3$-simplices are vertices, edges, triangles, and tetrahedra, respectively. 

\begin{figure}[b]
\centering
 \includegraphics[width=0.9\textwidth]{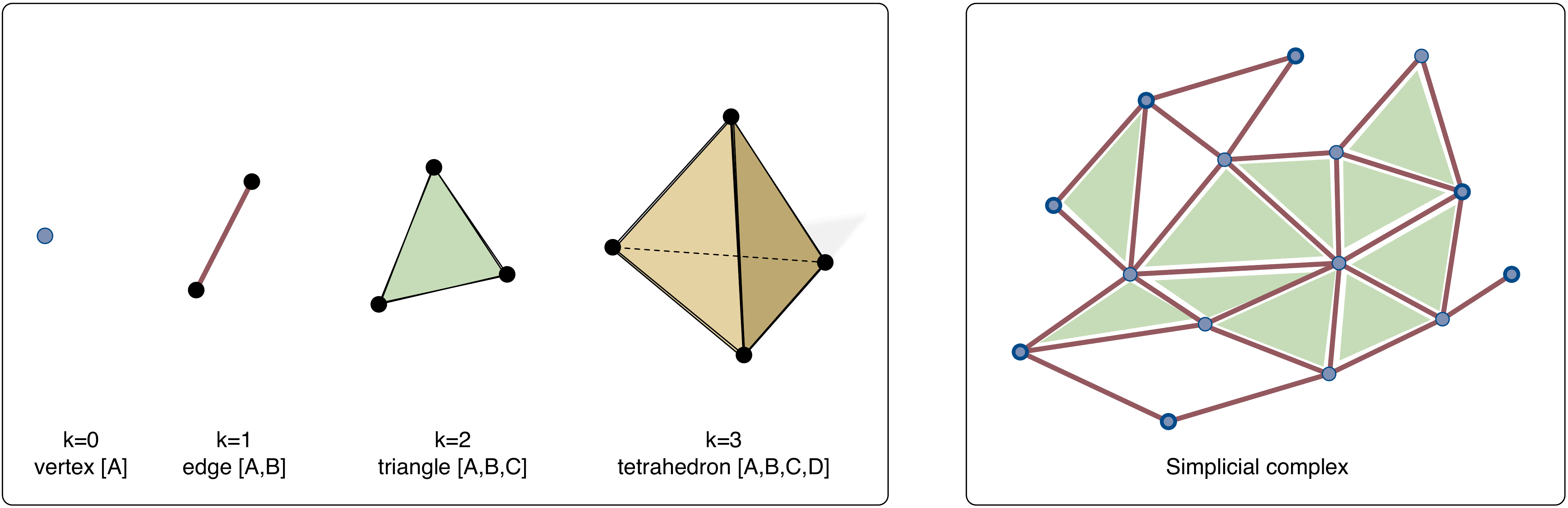}
\caption{Left: $0$-, $1$, $2$ and $3$-simplex, respectively. Right:  a simplicial complex embedded in $\Rspace^2$.}
\label{fig:simplicial}
\end{figure}

A \emph{simplicial complex} $K$ is a set of simplices such that every face  of a simplex from $K$ is also in $K$, and the intersections of two simplices in $K$ is either empty or a face of both simplices, see Fig.~\ref{fig:simplicial}.
A more detailed discussion can be found in~\cite{Munkres1984,Edelsbrunner2006}.
A simplicial complex is a type of \emph{cell complex} in which the cells are simplices. 
There are several different ways to formalize and instantiate the notion of a cell complex, including CW complex, $\Delta$-complex, cube complex, polytopal complex, etc.; see Hatcher~\cite{Hatcher2002} for an introduction. 

\subsubsection*{Neighborhood graphs}
Neighborhood graphs impose combinatorial structures on point clouds that capture certain notion of proximity. 
Such structures give rise to the use of grid-based analysis methods but are also of interest for clustering algorithms and many discrete theories. 
The most fundamental neighborhood structure is the Delaunay triangulation of a point cloud. 
Given a finite set of points $P=\{p_i\}_{i=1}^{m} \subset \Rspace^n$,  
the \emph{Voronoi diagram}\index{Voronoi diagram} is defined as a decomposition of the domain in regions $V_i$ assigned to each point $p_i \in P$. $V_i$ contains all points in $\Rspace^n$ that are at least as close to $p_i$ as to any other point in $P$.
The dual structure of the Voronoi diagram in the plane is the \emph{Delaunay triangulation} and in three dimension the \emph{Delaunay tetrahedralization}\index{Delaunay triangulation}\index{Voronoi diagram}. 
The Delaunay triangulation maximizes the minimum angle in a triangulation and gives rise to a reasonably nice triangulation.  
The concept extends to higher dimensions, but its computation becomes very costly. 
Many other neighborhood graphs have been studied with respect to geometric properties and robustness. Examples include the Gabriel graph~\cite{Gabriel1969} and the $k$-nearest neighbors graph.
A more detailed discussion about such graphs can be found in textbooks on computational geometry~\cite{DeBerg1997}. Neighborhood graphs in the context of high-dimensional and sparse data in visualization applications are also discussed in~\cite{Correa2011a}.
There is a large body of work related to meshing that is also relevant in this context~\cite{ZhangYJ2016}.

\subsubsection*{Reconstruction and interpolation}
The goal of a \emph{reconstruction} is to recover an approximate version of a continuous function from a sampled data set. A reconstruction that matches the values in the sampled points exactly is called \emph{interpolation}.

Given a set of points (vertices or nodes)    
    $P=\{p_i\}_{i=1}^{m}$ 
    with  $p_i \neq p_j$   for  $i \neq j$
and a set of associated values 
    $\{f_i \in\Rspace\}_{i=1}^{m}$, 
 a function $f:\Rspace^n\rightarrow \Rspace$ is called \emph{interpolating function} for the set of points if it fulfills  the interpolation condition
    $f(p_i) = f_i$, for $1 \leq i \leq m.$
    
Infinitely many possibilities are available to interpolate a set of points. The choice of a specific interpolation is often guided by simplicity and efficiency. 
It is important to be aware that different interpolation schemes may have significant impact on the computation and visualization results. 
The most common interpolation methods for gridded data are piecewise linear, bilinear, and trilinear interpolations. For scattered data, one typically constructs a grid or uses radial basis functions~\cite{Buhmann2003}. 

\subsubsection*{Discrete theories}
Discrete theories typically inherit structural properties from the smooth setting and come with theoretical understandings about the preservation of relevant invariants. 
In general, they satisfy a subset of properties from the smooth setting, resulting in a large diversity of discrete theories~\cite{Wardetzky2007}. For example, in the discrete setting, a geodesic defined as a locally shortest connection is not equivalent to the straightest connection, as in the continuous setting~\cite{Polthier1998}.

In visualization, the most important examples arise from combinatorial differential topology and geometry. 
For instance, discrete exterior calculus provides discrete differential operators~\cite{Desbrun2005a}; and discrete differential geometry introduces concepts for curvatures and geodesics~\cite{Desbrun2006}. A very useful and popular discrete theory is discrete Morse theory~\cite{Forman2001}, which forms the base of many current algorithms for the extraction of the Morse-Smale complex, see also Section~\ref{sec:topology}.

\section{Symmetries, Invariances, and Features}
\label{sec:features}
\setlength{\epigraphwidth}{0.8\textwidth}
\epigraph{Symmetry is a vast subject, significant in art and nature. Mathematics lies at its root, and it would be hard to find a better one on which to demonstrate the working of the mathematical intellect.}
{\textit{Hermann Weyl, mathematician and theoretical physicist~\cite{Weyl1952}}}
Symmetries, invariances, and conserved quantities are closely related concepts that play an important role in many mathematical and  physical theories, for instance, Noether's theorem links symmetries of physical spaces with conservation properties~\cite{Schwichtenberg2017}. 
Invariants are properties of an object (a system or a data set) that remain unchanged when certain transformations (such as rotations or permutations) are applied to the object. 
In visualization, invariants play a central role for feature definition and pattern recognition. 
For example, the number of legs of a 3-dimensional animal model is invariant with respect to changes due to animal movement or shape morphing. 
Another example is the Galilean invariance for flow features, e.g., vorticity does not change under certain changes of the reference frame even though the flow components change~\cite{Peacock2015}. 
There are also topological invariants which characterize spaces with respect to smooth deformations~\cite{Hatcher2002}.
A formal analysis of the symmetries that arise from group actions, with a strong emphasis on the geometry, Lie groups, and Lie algebra, can be found in textbooks dealing with representation theory and invariant theory~\cite{Goodman2009}.

\subsubsection*{Features, traits, and properties}
According to the Cambridge Dictionary, a feature is ``a typical quality or an important part of something".
In the visualization literature, the term feature is not well-defined and oftentimes an overloaded concept. 
Features often represent structures in a data set that are meaningful within some domain-specific context. 
They can be used as the basis for abstract visualization.
Here we define a feature $F(\Dcal) \subset \mathcal{I}$ of a data set $\Dcal$ as a subset of data items having a specific property; see Section~\ref{sec:data}.  
For field data, features are typically defined as certain subsets in the spatial domain. 
Typical features of a scalar field $s:D\rightarrow \Rspace$ are iso-surfaces $s^{-1}(a)$ (for $a \in \Rspace$), and the set of critical points of $s$.

In many cases, features can be locally defined by \emph{traits} $T \subset \mathcal{A^*}$, subsets of the enriched attribute space $\Acal^*$ containing the data attributes and possibly derived quantities.   
Specifically, given a field $f: D \to \Acal^*$ that maps a domain $D$ into an enriched attribute space $\Acal^*$, a \emph{trait-induced feature} is defined to be $F_T(\Dcal)=f^{-1}(T)=\{x\in \mathcal{I}\lvert f(x)\in T\}$, for some $T \subset \Acal^*$~\cite{Jankowai2018}. 
A point trait $T=\{p\} \in \Acal^*=\Rspace$ gives rise to a trait-induced feature known as an \emph{iso-surface}. 
A point trait is also referred to as a \emph{feature descriptor}. 
If $\Acal^*$ encodes the derivatives of $f$, then the set of critical points is a trait-induced feature given by all points where the derivative of the scalar function is equal to $0$. 
A line trait is a line in $\Acal^* = \Rspace^2$ spanned by the scalar values and its derivatives.  
It is desirable for a descriptor to be invariant with respect to changes (e.g.,~rotations and scalings) to the data representation. 

Other types of features based on structures of the data, such as cycles in a graph, may not be described by traits naturally.  
Such features are referred to as \emph{structure-induced features}. 
In general, features can be defined by any combination of attribute and structural constraints.

\subsubsection*{Transformations, symmetries, and invariances}
Invariants are directly linked to transformations $T$ describing an inherent \emph{symmetry}\index{symmetry} of the system.
A \emph{transformation}\index{transformation} is a function that maps a set $X$ to itself, i.e. $T : X \rightarrow X$.
In the context of visualization, a transformation concerning the structure $\Scal$ is called the \emph{inner transformation}\index{inner transformation}; a transformation of the attribute space $\Acal^*$ is  called the \emph{outer transformation}\index{outer transformation}.
A transformation can be both an inner and an outer transformation.
The notion of invariance and transformation can also be extended to  changes in the model used to create the visualization, or the image itself~\cite{Kindlmann2014}. 

When talking about invariants, we are interested not only in one specific transformation but also in certain classes of transformation described as transformation groups~\cite{Goodman2009}. 
A transformation group acting on a set $X$ is defined as a group $G$  with neutral element $e$ and an action 
\[T: G\times X \rightarrow X,\]
 where each group element $g\in G$ defines a transformation $T_g$ as  $T_g(x)\equiv T(g,x)$ 
with the following properties: for all $x\in X$ and all $g,h\in G$,                                      
$T_e(x) = x$ and
$T_g (T_h(x))=T_{gh}(x)$.

A \emph{symmetry group}\index{symmetry group} is a group that conserves a certain structure, property or feature. 
It gives a unique relation between symmetries  and invariants.  
Formally, let $T:\Dcal\rightarrow \Dcal$ be a transformation (short for  $T:\mathcal{I}\times\Acal^*\rightarrow \mathcal{I}\times\Acal^*$),
and $F(\Dcal)$ be a feature of a data set $\Dcal$. Then we say that $T$ is a \emph{symmetry}\index{symmetry} of $\Dcal$ if $F(\Dcal)$ commutes with the transformation $T$
\[
        T(F(\Dcal))=F(T(\Dcal)).
\]
Typical transformations for field data are rotations in 3-dimensional Euclidean space that form the group $SO_3$ acting on $\Rspace^3$. An application is the definition of invariant moments as descriptors of flow patterns~\cite{Bujack2014}.
An example that plays an important role in flow visualization is the Galilean transformation, which transforms coordinates between two reference frames that differ only by constant relative motion~\cite{Kasten2016}. 
Domain-specific invariants like shear stress or anisotropy also play a central role in tensor field visualization~\cite{Kratz2011}.  
An example of discrete data is the permutation group $Sym(M)$ whose elements are permutations of a set $M$.

\section{Cluster Analysis}
\label{sec:clusters}
\setlength{\epigraphwidth}{0.8\textwidth}
\epigraph{The Milky Way is nothing else but a mass of innumerable stars planted together in clusters.}
{\textit{Galileo Galilei, astronomer, physicist and engineer}}
\noindent
A frequently employed approach in visualization and exploratory analysis is cluster analysis or clustering, i.e., to assign a set of objects to groups in a manner such that objects in the same group are more similar to each other in some manner than to those in other groups. In other words, data are decomposed into a set of classes that in some sense reflect the distribution of the data.

To achieve this general goal, a very large variety of algorithms have been presented for specific problems or data modalities~\cite{Jain1999,Estivill2002}; they differ significantly in how they define and identify clusters. Clustering results are typically subject to various parameters, and it is often necessary to modify (e.g., transform) input data and choose parameters to obtain a result with desired properties. We describe four clustering techniques that are frequently applied in data analysis and visualization and illustrate how they have been used to address various visualization problems.

\para{$k$-means Clustering} Given a set of data $(x_1,\dots,x_n)$ where each $x_i$ is a $d$-dimensional real vector, $k$-means clustering (also called \emph{Lloyd's algorithm}) seeks to partition the data into $k \leq n$ disjoint sets $C = \{C_1,\dots,C_k\}$ (with a fixed $k$) such that the variance within each cluster is minimized, i.e., to find 
\[
    \underset{C}{\operatorname{arg\,min}}\ \sum_{i=1}^k\, \sum_{x \in C_i}\, \| x - m_i \|^2
\] 
where $m_i$ is the mean of data in $C_i$. The result depends centrally on the chosen metric, for which the Euclidean norm is often selected. Algorithmically, $C_i$ can be found iteratively in a manner similar to computing a centroidal Voronoi tessellation~\cite{Du1999}: given an initial set of cluster centers $m_i$, assign to each cluster the data points that are closer to $m_i$ than to all other cluster centers. 
Compute a new set of means as cluster centers from the assigned points, and repeat the process until convergence. Initially, the data centers can either be chosen randomly or according to heuristics~\cite{Celebi2013}. 

$k$-means clustering was used in visualization, for example, by Woodring and Shen~\cite{Woodring2009}, who employed it to automatically generate transfer functions for volume rendering temporal data. They achieved this by identifying clusters of data points that behave similarly over time. $k$-means clustering is relatively easy to understand and utilize. However, a major drawback of this approach is that the number  of classes or clusters $k$ must be specified a priori.

\para{Spectral clustering} Clustering is not directly applied on the data, but rather on the \emph{similarity matrix} $S$ (where $S_{ij} = \| x_i - x_j \|$) that contains pairwise distances between individual data items. Clustering is then performed on the eigenvectors of $S$. Intuitively, $S$ can be viewed as describing a mass-spring system. Masses coupled through tight springs will largely move together relative to the equilibrium of such a system, and thus eigenvectors of small eigenvalues of $S$ can be seen to form a suitable partition of the data.

As with clustering in general, many incarnations of this basic idea have been given. The \emph{normalized cuts} technique is a non-parametric clustering approach often used in image segmentation~\cite{Shi2000}. For visualization purposes, it was utilized by Ip et al.~to explore feature segmentation of three-dimensional intensity fields~\cite{Ip2012}, and by Brun et al. to visualize white matter fiber traces in DT-MRI data~\cite{Brun2004}.

\para{Density-based clustering} The \textsc{DBSCAN} (density-based spatial clustering of applications with noise) algorithm is a widely used general-purpose clustering scheme~\cite{Ester1996,Schubert2017}. It considers the density of data points in their embedding space and subdivides them into three types. A point $x_i$ is a \emph{core points} if at least $m$ points lie within a distance of $\delta$ from $x_i$; these  points are called \emph{directly reachable} from $x_i$. Both $m$ and $\delta$ are parameters. An arbitrary point $x_j$ is \emph{reachable} from $x_i$ if there is a path $x_i,x_k,\dots,x_j$ such that each point in the path is directly reachable from its predecessor. Points that are not reachable from any core point are called \emph{outliers}. Clusters are formed by core points and the points that are reachable from them. (There may be multiple core points in a cluster.) Due to the non-symmetric reachability relations, DBSCAN uses the notion of density-connectedness for a pair $x_i$ and $x_j$. 
That is, points $x_i$ and $x_j$ are connected if there is a third point $x_l$ from which both $x_i$ and $x_j$ are reachable. 

DBSCAN is relatively easy to implement and has good runtime properties, but many variants of the basic technique exist that differ in various details~\cite{Suthar2013,Schubert2017}. Wu et al. used DBSCAN to provide level-of-detail in visualization and exploration of academic career path~\cite{Wu2013}.

\para{Mean shift} A \emph{mean shift} procedure is a variant of density-based clustering; it is applied to identify the maxima (or \emph{modes}) of a density function from discrete samples. 
Fixing a kernel function $K(x_i - x)$ (typically flat or Gaussian) and a point $x$ in the embedding space, the weighted mean in a window around $x$ is 
\[
m(x) = \frac{\sum_{x_i} K(x_i - x)x_i}{\sum_{x_i} K(x_i - x)}.
\]
The mean shift $m(x) - x$ is then minimized by setting $x \leftarrow m(x)$ and iterating until convergence. Data points $x_j$ are grouped into clusters according to the mode to which the mean shift converges if initialized with $x_j$. This process yields a general-purpose clustering technique that does not incorporate assumptions about the data and relies on a single parameter, the kernel bandwidth. In visualization, a good example of the usefulness of this algorithm is given by B\"ottger et al.~\cite{Bottger2013}, who use mean-shift clustering to achieve edge bundling in brain functional connectivity graphs.

\section{Statistics for Visualization}
\label{sec:statistics}
\setlength{\epigraphwidth}{0.8\textwidth}
\epigraph{If the statistics are boring, you've got the wrong numbers. }
{\textit{Edward R. Tufte, statistician~\cite{Tufte2001}}}
\noindent
Statistics deals with the collection, description, analysis and interpretation of (data) \emph{populations}. \emph{Descriptive statistics} are used to summarize population data. \emph{Moments}, also called \emph{summary statistics}, are a statistical notion to describe the shape of a function (distribution). Mathematically, the $n$-th \emph{central moment} of a real-valued continuous function $f(x)$ of a real variable is given by 
\[
    \mu_n\ =\ \int_{-\infty}^{\infty} (x-c)^n f(x) dx,
\]
where $c$ is the mean of $f(x)$. The first moment corresponds to the mean, and a usual assumption considers $c=0$. These moments give rise to the usual statistical descriptors of a distribution such as variance ($n=2$), skewness ($n=3$), and kurtosis ($n=4$). Potter et al. provide guidance on the visualization of functions via their summary statistics~\cite{Potter2010}. For multiple variables, the concept of moments can be generalized to \emph{mixed moments}. Applications in visualization include pattern matching for feature extraction~\cite{Bujack:2018:TVCG}. 

A frequent problem in comparative visualization is comparing distributions. Here, the \emph{covariance} of two distributions
\[
\operatorname{cov}(f,g)\ =\ E\left[f-E[f]\right]\,E\left[g - E[g]\right]
\]
signifies their joint variability. 
In the multivariate case, covariance can be generalized to the \emph{covariance matrix}. Covariance matrices have been frequently used in visualization, for example in glyph-based~\cite{Post1995} or feature-based visualization~\cite{Wong2000}.

Furthermore, \emph{correlation} of functions may be used for comparison. In the broadest sense, correlation is any statistical association between data populations; in practice, correlation is usually used to indicate a linear relationship between functions. An commonly used concept is the \emph{Pearson's correlation coefficient},
\[
\rho_{f,g}\ =\ \operatorname{corr}(f,g)\ =\ \frac{\operatorname{cov}(f,g)}{\sigma_f \sigma_g},
\]
where $\sigma_f$ and $\sigma_g$ refer to the standard deviation of $f$ and $g$, respectively. 
$\rho_{f,g} \in (0,1]$ if $f$ and $g$ are positively correlated; 
$\rho_{f,g} \in [-1,0)$ if $f$ and $g$ are negatively correlated; 
$\rho_{f,g} = 0$ if $f$ and $g$ have no linear correlation. 
Finding correlations among data is one of the most essential tasks in many scientific problems, and visualization can be very helpful during such a process~\cite{Chen2011,Gosink2007}.

\emph{Order statistics}, on the other hand, characterizes a population in terms of ordering and allows us to make statistical statements about the distribution of its values.
For example, the $q$-percentile ($0 \leq q \leq 100$) denotes the value below which $q$ percent of the samples are located. Order statistics can be easily combined with descriptive statistics in the univariate case~\cite{Potter2010}. Higher dimensional variants of these notions are also available and used to represent data visually~\cite{Raj2018}.
An interesting generalization of order statistics to a widely-used topological structure is the contour boxplot~\cite{Whitaker2013}.

\section{Topological Data Analysis}
\label{sec:topology}
\setlength{\epigraphwidth}{0.8\textwidth}
\epigraph{If you can put it on a necklace, it has a one-dimensional hole. If you can fill it with toothpaste, it has a two-dimensional hole. For holes of higher dimensions, you are on your own.
}{\textit{Evelyn Lamb, math and science writer~\cite{Lamb2014}}}
\noindent
For topology in visualization, two key developments from computational topology play an essential role in connecting mathematical theories to practice: 
first, separating features from noise using \emph{persistent homology}; 
second, abstracting topological summaries of data using \emph{topological structures} such as Reeb graphs, Morse-Small complexes, Jacobi sets, and their variants. 

\subsubsection*{Topology, homology and Betti numbers}
Topology has been one of the most exciting research fields in modern mathematics~\cite{James1999}. 
It is concerned with the properties of space that are preserved under continuous deformations, such as stretching, crumpling, and bending, but not tearing or gluing~\cite{WikipediaTopology2018}. 

The beginning of topology was arguably marked by Leonhard Euler, who published a paper in 1736 that solved the now famous K\"{o}nigsberg bridge problem. 
In the paper, titled \emph{``The Solution of a Problem Relating to the Geometry of Position"}, Euler was dealing with ``a different type of geometry where distance was not relevant"~\cite{OConnorRobertson1996}.  
Johann Benedict Listing was credited as the first to use the word ``topology" in print based on his 1847 work titled \emph{``Introductory Studies in Topology"}; although many of Listing's topological ideas were borrowed from Carl Friedrich Gauss~\cite{OConnorRobertson1996}.  
Both Listing and Bernhard Riemann studied the \emph{components} and \emph{connectivity} of surfaces. 
Listing examined connectivity in $3$-dimensional Euclidean space, and Enrico Betti extended the idea to $n$ dimensions. 
Henri Poincar\'{e} then gave a rigorous basis to the idea of connectivity in a series of papers \emph{``Analysis situs"} in 1895. He introduced the concept of \emph{homology} and improved upon the precise definition of Betti numbers of a space~\cite{OConnorRobertson1996}. 
In other words, it was Poincar\'{e} who ``gave topology wings"~\cite{James1999} via the notion of homology. 

\begin{figure}[t]
    \begin{center}
 \includegraphics[width=1\linewidth]{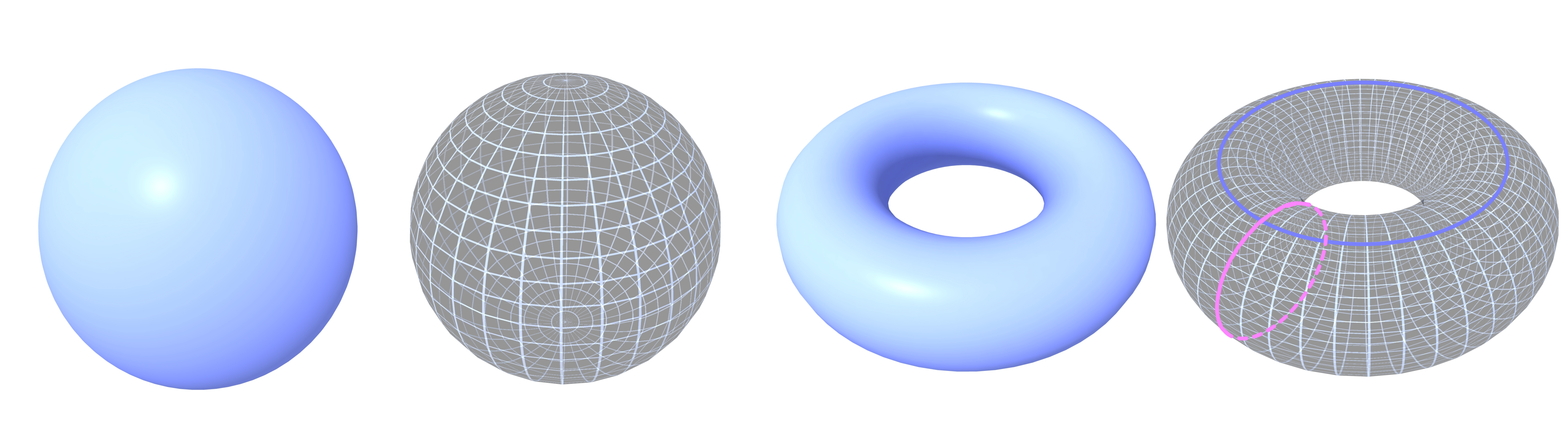} 
        \vspace{-4mm}
        \caption{Betti numbers for the sphere and the torus. $\beta_0 = 1$, $\beta_1 = 0$, and $\beta_2 = 1$ for the sphere (left) and $\beta_0 = 1$, $\beta_1 = 2$, and $\beta_2 = 1$ for the torus (right). Image courtesy of Mustafa Hajij.}
        \label{fig:Betti}
    \end{center}
\end{figure}

The original motivation to define homology was that it can be used to tell two objects (a.k.a. topological spaces) apart by examining their holes. 
This process associates a topological space $\Xspace$ with a sequence of abelian groups called homology groups $\Hgroup(\Xspace)$, which, roughly speaking, count and collate \emph{holes} in a space~\cite{Ghrist2008}. 
Informally, homology groups generalize a common-sense notion of connectivity. 
They detect and describe the connected components ($0$-dimensional holes), tunnels ($1$-dimensional holes), voids ($2$-dimensional holes), and holes of higher dimensions in the space. 
The $p$-th Betti number $\beta_p$ is the rank of the $p$-th homology group of $\Xspace$, $\Hgroup_p(\Xspace)$, and captures the number of $p$-dimensional holes of a topological space. 
For instance, a sphere contains no tunnels but a void, and a torus contains two tunnels (see Fig.~\ref{fig:Betti}).

\subsubsection*{From homology to persistent homology}
For simplicity, we work with data represented by simplicial complexes denoted by $\Xspace$. 
In algebraic terms, the construction of homology groups
begins with a chain complex $\Cgroup(\Xspace)$ that encodes information about $\Xspace$, which is a sequence of abelian groups
$\Cgroup_0(\Xspace), \Cgroup_1(\Xspace), \dots$ connected by homomorphisms known as the boundary operators $\bdr_{p} : \Cgroup_p(\Xspace) \to \Cgroup_{p-1}(\Xspace)$. 
The \emph{$p$-th homology group} is defined as $\Hgroup_p(\Xspace) = \myker (\bdr_p)/ \myim (\bdr_{p+1})$.
The \emph{$p$-th Betti number} is the rank of this group, $\beta_p = \rank \Hgroup_p$, see~\cite{Munkres1984} for an introduction. 

\begin{figure}[t]
    \begin{center}
\includegraphics[width=1\linewidth]{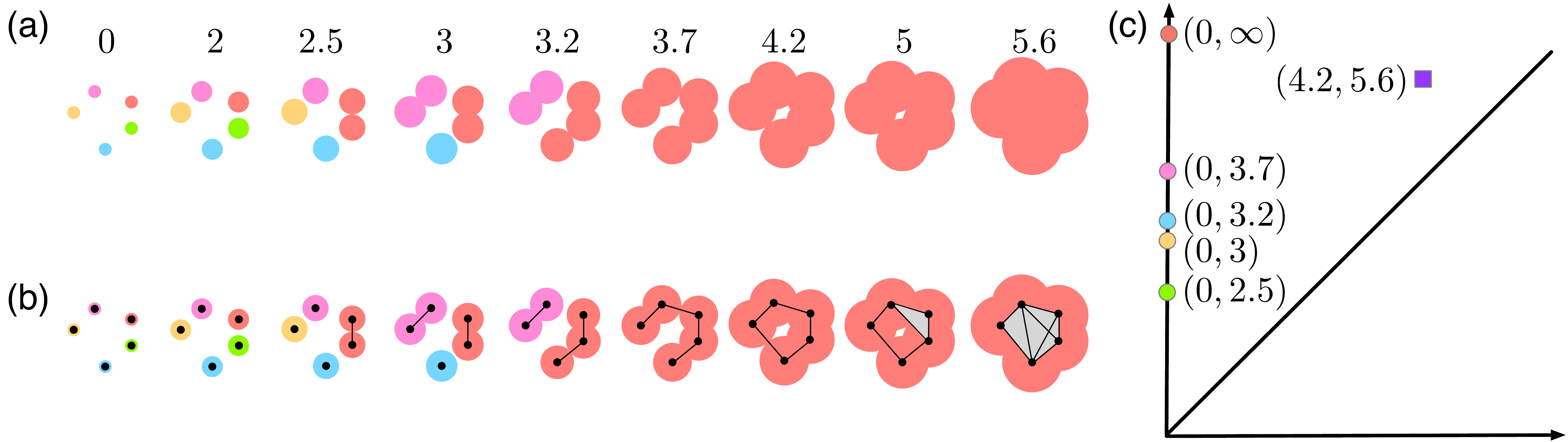} 
        \vspace{-2mm}
        \caption{Computing persistent homology of a point cloud in $\Rspace^2$. (a) A nested sequence of topological spaces formed by unions of balls at increasing parameter values. (b) A filtration of simplicial complexes that capture the same topological information as in (b). (c) $0$- (circles) and $1$-dimensional (squares) features in a persistence diagram.}
        \label{fig:persistence}
    \end{center}
\end{figure}

Persistent homology transforms the algebraic concept of homology into a multi-scale notion by constructing an extended series of homology groups. 
In its simplest form, persistent homology applies a homology functor to a sequence of topological spaces connected by inclusions, called a \emph{filtration}. 
Consider a finite sequence of simplicial complexes connected by inclusions $f_p^{i,j}: \Xspace_i \xhookrightarrow{} \Xspace_j$, 
$$\emptyset = \Xspace_0 \xhookrightarrow{} \Xspace_1 \xhookrightarrow{}  \cdots \xhookrightarrow{} \Xspace_n = \Xspace.$$
Applying $p$-th homology to this sequence results in a sequence of homology groups connected from left to right by homomorphisms induced by the inclusions, 
$$0 = \Hgroup_p(\Xspace_0) \to \Hgroup_p(\Xspace_1) \to \cdots \to \Hgroup_p(\Xspace_n) =  \Hgroup_p(\Xspace) $$ 
for each dimension $p$.  
The \emph{$p$-th persistent homology group} is the image of the homomorphism induced by inclusion, $\Hgroup_p^{i,j} = \myim f_{p}^{i,j}$ for $0 \leq i \leq j \leq n$. 
The corresponding \emph{$p$-th persistent Betti number} is the rank of this group, $\beta_p^{i,j} = \rank \Hgroup_p^{i,j}$~\cite[Page 151]{EdelsbrunnerHarer2010}.  
As the index increases, the rank of the homology groups changes. 
When the rank increases (i.e.,~$f_p^{i-1, i}$ is not surjective), we call this a \emph{birth} event at $\Xspace_i$; when the rank desreases (i.e.,~$f_p^{j-1, j}$ is not injective), we call this a \emph{death} event at $\Xspace_j$. 
Persistent homology pairs the birth and the death events as a multi-set of points in the plane called the \emph{persistence diagrams}~\cite{EdelsbrunnerLetscherZomorodian2002}; see~\cite{EdelsbrunnerMorozov2012, EdelsbrunnerMorozov2017} for a comprehensive mathematical introduction.   
A celebrated theorem of persistent homology is the \emph{stability}
of persistence diagrams~\cite{Cohen-SteinerEdelsbrunnerHarer2007}, that is, small changes in the data lead to small changes in the corresponding diagrams, making it suitable for robust data analysis.   
See Fig.~\ref{fig:persistence} for an example in $\Rspace^2$. 
Given a set of points in $\Rspace^2$, we compute its persistent homology by studying the union of balls centered around the points as the radius increases. 
Here, a green component is born at time $0$ and dies when it merges with a red component at time $2.5$, resulting a point $(0, 2.5)$ in the persistence diagram. 
A tunnel is born at time $4.2$ and dies at time $5.6$, giving rise to a point $(4.2, 5.6)$ in the persistence diagram. 

\subsubsection*{Topological structures}
\begin{figure}[t]
    \begin{center}
\includegraphics[width=0.8\linewidth]{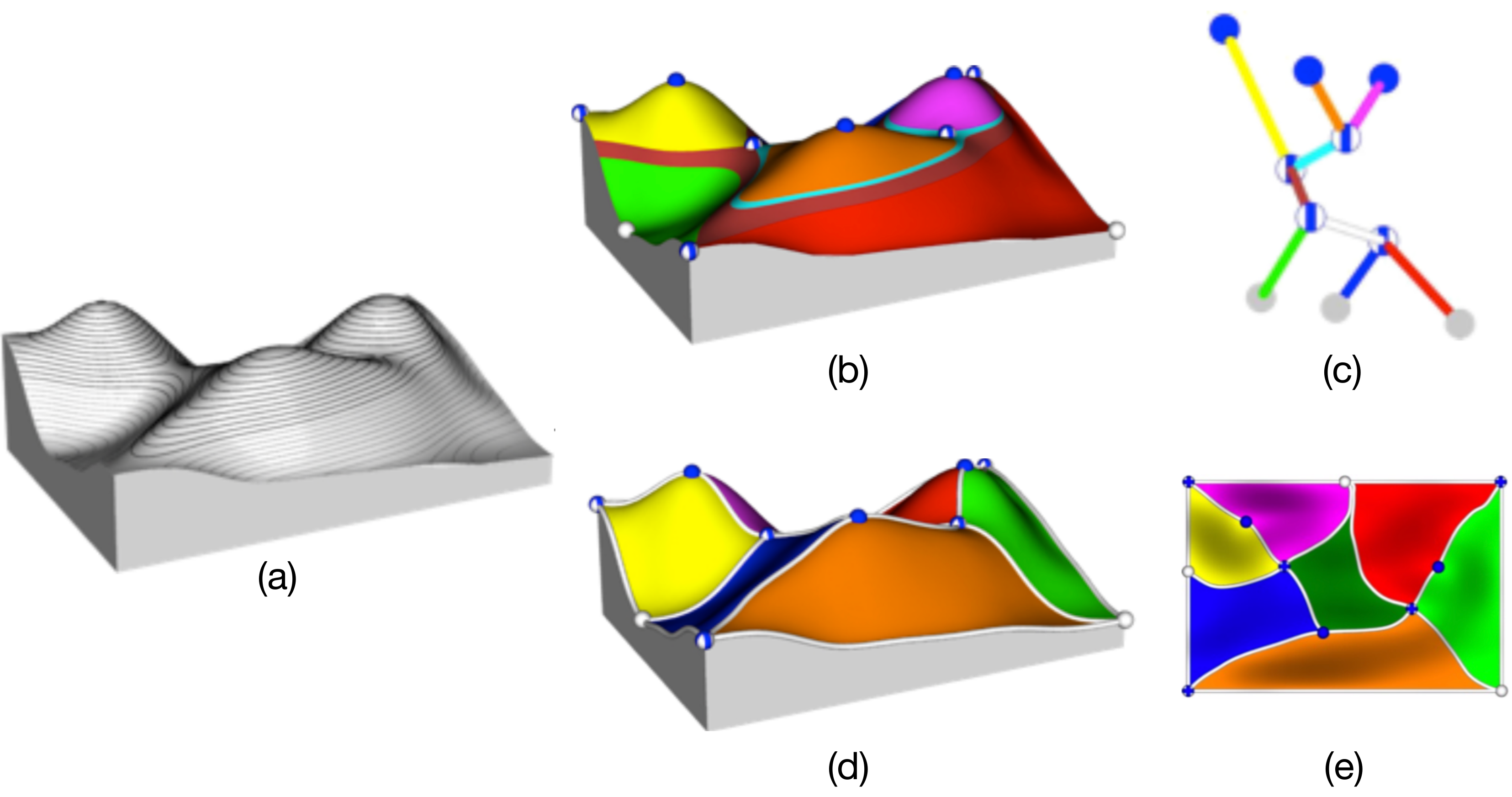} 
        \vspace{-2mm}
        \caption{Contour-based (c) and gradient-based (e) topological structures of a $2$-dimensional scalar function (a).}
        \label{fig:topostructure}
    \end{center}
\end{figure}

Several techniques in topological data analysis and visualization construct topological structures from well-behaved functions on point clouds as summaries of data. 
On one hand, the well-behave-ness is formalized with the Morse theory. 
On the other hand, such topological structures can be roughly classified into two types: contour-based (Reeb graphs~\cite{Reeb1946}, Reeb spaces~\cite{EdelsbrunnerHarerPatel2008}, contour trees~\cite{CarrSnoeyinkAxen2003} and merge trees), and gradient-based topological structures (Morse-Smale complexes~\cite{EdelsbrunnerHarerNatarajan2003,EdelsbrunnerHarerZomorodian2003} and Jacobi sets~\cite{EdelsbrunnerHarer2002}), see Fig.~\ref{fig:topostructure}. 
All such topological structures provide meaningful abstractions of (potentially high-dimensional) data, reduce the amount of data needed to be processed or stored, utilize sophisticated hierarchical representations that capture features at multiple scales, and enable progressive simplifications~\cite{LiuMaljovecWang2017}.

\para{Morse function} 
Let $\Mspace$ be a smooth, compact, and orientable $d$-manifold
without boundary ($d \geq 2$). 
Suppose $\Mspace$ is equipped with a Riemannian metric so that gradients are well defined. 
Given a smooth function $f: \Mspace \to \Rspace$, a point $x \in \Mspace$ is
called a \emph{critical point} if the gradient of $f$ at $x$ equals zero, that is, 
$\grad f(x) = 0$, and the value of $f$ at $x$ is called a \emph{critical
value}. All other points are \emph{regular points} with their function values
being \emph{regular values}. A critical point is \emph{non-degenerate} if the
Hessian, i.e., the matrix of second partial derivatives at the point, is
invertible.  
A smooth function $f$ is a \emph{Morse function} if (a) all its critical
points are non-degenerate; and (b) all its critical values are distinct~\cite[Page 128]{EdelsbrunnerHarer2010}. 
A pair of two Morse functions is \emph{generic} if their critical points do not overlap. 

\para{Morse-Smale complexes}
Given a Morse function $f: \Mspace \to \Rspace$, at any regular point $x$
 the gradient is well-defined and integrating it in both directions traces out
 an integral line, $\gamma: \Rspace \to \Mspace$, which is a maximal path whose
 tangent vectors agree with the gradient \cite{EdelsbrunnerHarerZomorodian2003}.
Each integral line begins and ends at critical points of $f$.  
The \emph{ascending/descending manifolds} of a
critical point $x$ is defined as all the points whose integral lines start/end
at $x$.  The descending manifolds form a complex called a \emph{Morse complex}
of $f$ and the ascending manifolds define the Morse complex of $-f$.  The set of
intersections of ascending and descending manifolds creates the
\emph{Morse-Smale} complex of $f$.  Each \emph{cell} of the
The Morse-Smale complex is a union of integral lines that all share the same origin
and the same destination.  
In other words, all the points inside a single cell have uniform gradient flow behavior. 
These cells yield a decomposition into monotonic,
non-overlapping regions of the domain, as shown in Fig.~\ref{fig:topostructure}(b) for a
$2$-dimensional height function.

\para{Jacobi set for a pair of Morse functions} 
Given a generic pair of Morse functions, $f, g: \Mspace \to \Rspace$, their Jacobi set $\J = \J(f,g) = \J(g,f)$ is the set of points where their gradients are 
      parallel or zero~\cite{EdelsbrunnerHarer2002}. That is, for some $\lambda \in \Rspace$, 
\begin{eqnarray}
\label{eq:align}
\J = \{x \in \Mspace \mid \grad f(x) + \lambda \grad g(x) = 0 \mbox{ or } \grad g(x) + \lambda \grad f(x) = 0\}.
\end{eqnarray}
The sign of $\lambda$ for each $x$ is called its \emph{alignment}, as
it defines whether the two gradients are aligned or anti-aligned.  By
definition, the Jacobi set contains the critical points of both $f$ and $g$.

There exist several other descriptions of Jacobi sets~\cite{EdelsbrunnerHarer2002,
 EdelsbrunnerHarerNatarajan2004, NagarajNatarajan2011}.  One particularly useful description
is in terms of the \emph{comparison measure}, $\kappa$~\cite{EdelsbrunnerHarerNatarajan2004},
which is a gradient-based metric to compare two functions. It plays a
significant role in assigning an importance value to subsets of a Jacobi set in terms
of the underlying functions $f$ and $g$ by measuring the relative orientation
of their gradients.

\para{Reeb graphs and contour trees}
Let $f: \Xspace \to \Rspace^d$ be a generic, continuous mapping. 
Two points $x, y \in \Xspace$ are \emph{equivalent}, demoted by $x \sim y$, if $f(x) = f(y)$ and $x$ and $y$ belong to the same path-connected component of the pre-image of $f$, $f^{-1}(f(x)) = f^{-1}(f(y))$.
The \emph{Reeb space}, $\Rcal(X, f) = \Xspace / \sim$, is the quotient space contained by identifying equivalent points together with the quotient topology inherited from $\Xspace$. 
A powerful analysis tool, the \emph{Reeb graph}, is  a special case when $d = 1$. 

The Reeb graph of a real-valued function $f: \Xspace \to \Rspace$ describes the connectivity of its level sets. 
A \emph{contour tree} is a special case of the Reeb
graph if the domain $\Xspace$ is simply connected, see Fig.~\ref{fig:topostructure}(c). 
A \emph{merge tree} is similar to the Reeb graphs
and contour trees except that it describes the connectivity of
sublevel sets rather than level sets.
The Reeb graph stores information regarding the number of components at any
function value as well as how these components split and
merge as the function value changes. Such an abstraction
offers a global summary of the topology of the level sets and connects naturally with visualization. 

\section{Color spaces}
\par
\setlength{\epigraphwidth}{0.8\textwidth}
\epigraph{Although many great thinkers have held that an analytical or mathematical treatment of the subject is impossible or even undesirable, they have gradually deserted the field so that today and indeed throughout the past 50 years it has been generally recognized that a theory of color perception must be, both in form and content, a mathematical theory.
}{\textit{Howard L. Resnikoff, mathematician and business executive~\cite{resnikoff1974differential}}}

\noindent
Color is one of the central aspects of visualization and against common belief, a surprisingly mathematical one.
Operations on color are an important aspect in many applications, e.g., color mapping, re-sampling of color images or movies, and image manipulations, such as stitching, morphing, or contrast adaption. These operations can be expressed through mathematical formulae if the colors themselves can be expressed as elements of a mathematical space, in which certain concepts such as sums or distances have a meaning. 
However, as we will see, this is not easy. 

\begin{figure}[!ht]
\begin{minipage}[t]{0.3\linewidth}
\vspace{0pt}
\includegraphics[width=1\linewidth]{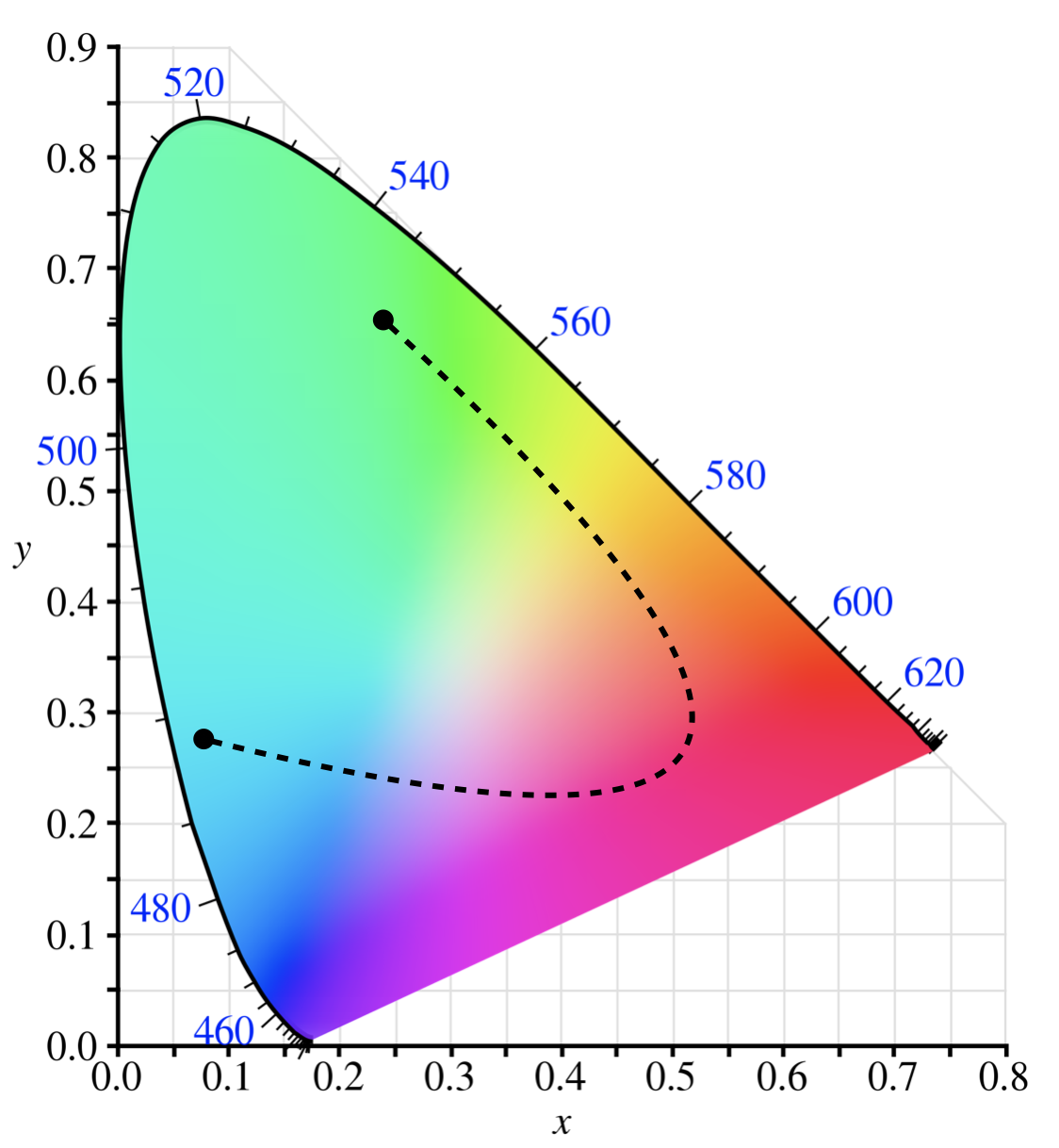}
\caption{CIE XYZ chromaticity diagram and a path that represents a colormap.} \label{fig:chromaticity}
\end{minipage}\hfill
  \begin{minipage}[t]{0.25\linewidth}
  \vspace{0pt}
   \includegraphics[width=1\linewidth]{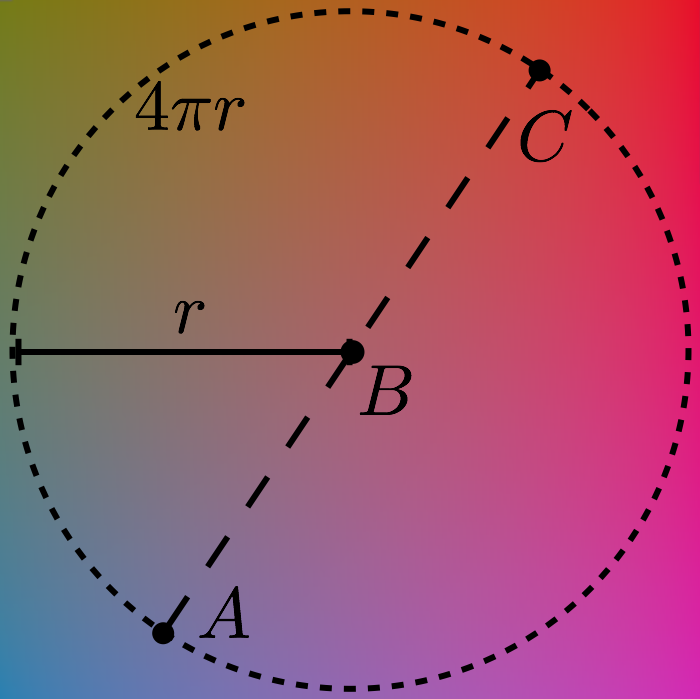}
    \caption{Illustration of hue-superimportance with circumference of $\approx 4 \pi r$ and diminishing returns ($\overline{AB} + \overline{BC} > \overline{AC})$.}\label{fig:principle}
  \end{minipage}\hfill
    \begin{minipage}[t]{0.35\linewidth}
  \vspace{0pt}
\includegraphics[width=0.8\linewidth]{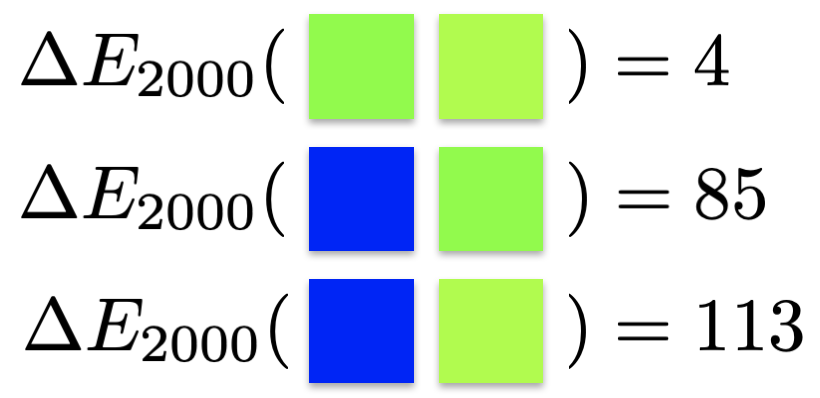}
        \caption{Illustration of non-metric behavior of CIE $\Delta E_{2000}$. Violation of the triangle inequality implies that the path over green RGB=(146,252,77) is shorter than the direct path from blue RGB=(0,0,255) to yellowish green RGB=(177,253,79), which is very counter intuitive.}\label{fig:nonMetric}
  \end{minipage}
\end{figure}

The space of all colors is in principal infinite-dimensional because any function over the frequencies of the visible spectrum forms a color. Since, however, the human eye has only three receptors for color, the space of distinguishable colors for humans is only three-dimensional~\cite{grassmann1853theorie,von1867handbuch}.
Depending on the choice of the three basis dimensions, many different colorspaces were developed. 
In displays, the basic colors are usually red, green, and blue (RGB) and for printing, the standard is cyan, magenta, yellow, and key black (CMYK).
The XYZ space by the Commission Internationale de L'Eclairage (CIE) is considered as the basis of all modern color spaces~\cite{guild1932colorimetric, international2004colorimetry}.
It embeds all visible colors unambiguously into one space of three imaginary primaries~\cite{fairman1997cie,broadbent2008calculation}. The chromaticity diagram in Fig.~\ref{fig:chromaticity} is the result of projecting XYZ to the Maxwell triangle $x+y+z=1$, which forms a representation of all visible hues and saturations.  

A number of spaces, e.g., CIELAB, CIELUV, and DIN99, CIECAM~\cite{international2004colorimetry, buring2002eigenschaften}, were defined as transformations of XYZ to derive
an ideal color space~\cite{judd1970ideal}, where the Euclidean distance is proportional to the perceived color difference. 

Human color perception has been known for a while to be non-Euclidean due to the principle called \emph{hue superimportance}~\cite{judd1968ideal} (cf. Fig.~\ref{fig:principle}). It refers to the fact that changes in hue are perceived more strongly than changes in saturation. The circumference of a circle of constant luminance and saturation would be estimated to measure about $4\pi$ for its radius, which cannot be embedded in a Euclidean plane.
Please note that the length $l\in\Rspace$ of a path $c:\Rspace\to C$ is defined for arbitrary metric spaces $C$ 
\begin{equation}
l=\sup_{0=t_0,...,t_n=1}\sum_{i=1}^k \Delta E (c(t_i),c(t_{i+1})).
\end{equation}
Therefore, classic descriptions of color spaces, such as those of von Helmholtz \cite{von1867handbuch}, Schr\"odinger~\cite{schrodinger1920grundlinien}, and Stiles~\cite{wyszecki1982color}, are based on Riemannian manifolds.

However, state-of-the-art research indicates that human color perception is also non-Riemannian, due to the further principle of \emph{diminishing returns}~\cite{judd1968ideal}, see Fig.~\ref{fig:principle}. In this context, diminishing returns refers to the phenomenon that when presented with two colors $A$ and $C$ and their perceived middle (average/mixture) $B$, an observer usually judges the sum of the perceived differences of each half greater than the difference of the two outer colors $\Delta (A,B) + \Delta (B,C) > \Delta (A,C)$.
This effect is produced by a natural contrast enhancement filter employed into the human perceptual system to adapt to different viewing conditions.
This property is dependent upon the distance between colors, especially for large distances.

As a result, modern color difference formulas (e.g.,~CIEDE1994, CIEDE2000) that were designed to match experimental data produce complicated spaces, which come with challenges. For example, they are not metric spaces. Being a metric is a very basic mathematical property that we would expect from a distance measure $d:C\times C\to\Rspace$, i.e.,~that it suffices non-negativity $d(x,y) \ge 0$, identity of indiscernible $d(x,y) = 0 \Leftrightarrow x = y$, symmetry $d(x,y)  = d(y,x)$, and the triangle inequality $d(x,z) \le d(x,y) + d(y, z)$. The reasons for such a challenge are not in the experimental data but can be found in the mathematical models underlying the distance formulae~\cite{mahy1994evaluation,luo2001development,huertas2006performance}. 
An example of the violation of the triangle inequality is shown in Fig.~\ref{fig:nonMetric}.

The difficulties, however, lie not only in the modeling of the color spaces but also in the visualization side. Mathematical operations on color become significantly harder in non-Euclidean spaces.
As a basic example, consider linear interpolation where values are taken equidistantly on a straight line connecting two points.
In non-Euclidean spaces, the concept of a straight line is, in general, undefined.

To overcome some of these difficulties, some authors generate spaces that are close to the original distance measure but are Euclidean or at least Riemannian~\cite{urban2007embedding, raj2012riemannian}.  This, however, conflicts with the experimental results from the perceptual sciences. We believe that future color spaces will continue to better approximate human color perception and embrace its complicated non-Euclidean structure because our computational capacities will enable us to work with them despite those difficulties. 
We believe that the path forward lies in improving visualization algorithms so that they run on general non-Euclidean color spaces. A few results  been obtained recently for color interpolation~\cite{zeyen2018interpolation} and colormap assessment~\cite{Bujack:2018:TVCG}.

\section*{Acknowledgement}
The authors would like to thank the organizers of Dagstuhl Seminar 18041 titled ``Foundations of Data Visualization" in 2018. 
BW is partially supported by NSF IIS-1910733, DBI-1661375, and IIS-1513616.
RB is partially supported by the Laboratory Directed Research and Development (LDRD) program of Los Alamos National Laboratory (LANL) under project number 20190143ER. 
IH is supported through Swedish e-Science Research Center (SeRC) and the ELLIIT environment for strategic research in Sweden. 
 


\begin{thebibliography}{100}

\bibitem{amann2011ordinary}
H.~Amann.
\newblock {\em Ordinary Differential Equations: An Introduction to Nonlinear
  Analysis}, volume~13 of {\em Studies in Mathematics}.
\newblock Walter de Gruyter, 2011.

\bibitem{Berger2003}
M.~Berger.
\newblock {\em {A Panoramic View of Riemannian Geometry}}.
\newblock Springer, 2003.

\bibitem{Bottger2013}
J.~{B{\"o}ttger}, A.~{Sch{\"a}fer}, G.~{Lohmann}, A.~{Villringer}, and D.~S.
  {Margulies}.
\newblock Three-dimensional mean-shift edge bundling for the visualization of
  functional connectivity in the brain.
\newblock {\em IEEE Transactions on Visualization and Computer Graphics},
  20(3):471--480, March 2014.

\bibitem{broadbent2008calculation}
A.~Broadbent.
\newblock Calculation from the original experimental data of the {CIE} 1931
  {RGB} standard observer spectral chromaticity coordinates and color matching
  functions.
\newblock {\em Qu{\'e}bec, Canada: D{\'e}partement de g{\'e}nie chimique,
  Universit{\'e} de Sherbrooke}, 2008.

\bibitem{Brun2004}
A.~Brun, H.~Knutsson, H.-J. Park, M.~E. Shenton, and C.-F. Westin.
\newblock Clustering fiber traces using normalized cuts.
\newblock In C.~Barillot, D.~R. Haynor, and P.~Hellier, editors, {\em Medical
  Image Computing and Computer-Assisted Intervention}, pages 368--375. Springer
  Berlin Heidelberg, 2004.

\bibitem{Buhmann2003}
M.~D. Buhmann.
\newblock {\em Radial Basis Functions: Theory and Implementations}.
\newblock Cambridge University Press, 2003.

\bibitem{Bujack2014}
R.~{Bujack}, I.~{Hotz}, G.~{Scheuermann}, and E.~{Hitzer}.
\newblock Moment invariants for {2D} flow fields via normalization in detail.
\newblock {\em IEEE Transactions on Visualization and Computer Graphics},
  21(8):916--929, Aug 2015.

\bibitem{Bujack:2018:TVCG}
R.~Bujack, T.~L. Turton, F.~Samsel, C.~Ware, D.~H. Rogers, and J.~Ahrens.
\newblock The good, the bad, and the ugly: A theoretical framework for the
  assessment of continuous colormaps.
\newblock {\em IEEE Transactions on Visualization and Computer Graphics},
  24(1):923--933, Jan 2018.

\bibitem{buring2002eigenschaften}
H.~B{\"u}ring.
\newblock Eigenschaften des farbenraumes nach din 6176 (din99-formel) und seine
  bedeutung f{\"u}r die industrielle anwendung.
\newblock In {\em Proceedings of 8th Workshop Farbbildverarbeitung der German
  Color Group}, pages 11--17, 2002.

\bibitem{butcher2016numerical}
J.~C. Butcher.
\newblock {\em Numerical Methods for Ordinary Differential Equations}.
\newblock John Wiley \& Sons, 2016.

\bibitem{CarrSnoeyinkAxen2003}
H.~Carr, J.~Snoeyink, and U.~Axen.
\newblock Computing contour trees in all dimensions.
\newblock {\em Computational Geometry}, 24(2):75--94, 2003.

\bibitem{Celebi2013}
M.~E. Celebi, H.~A. Kingravi, and P.~A. Vela.
\newblock A comparative study of efficient initialization methods for the
  k-means clustering algorithm.
\newblock {\em Expert Systems with Applications}, 40(1):200 -- 210, 2013.

\bibitem{Chen2011}
C.~{Chen}, C.~{Wang}, K.~{Ma}, and A.~T. {Wittenberg}.
\newblock Static correlation visualization for large time-varying volume data.
\newblock In {\em {IEEE} Pacific Visualization Symposium}, pages 27--34, 2011.

\bibitem{coddington2012introduction}
E.~A. Coddington.
\newblock {\em An introduction to ordinary differential equations}.
\newblock Courier Corporation, 2012.

\bibitem{coffin1911vector}
J.~G. Coffin.
\newblock {\em Vector analysis: An introduction to vector-methods and their
  various applications to physics and mathematics}.
\newblock J. Wiley \& Sons, 1911.

\bibitem{Cohen-SteinerEdelsbrunnerHarer2007}
D.~Cohen-Steiner, H.~Edelsbrunner, and J.~Harer.
\newblock Stability of persistence diagrams.
\newblock {\em Discrete and Computational Geometry}, 37(1):103--120, 2007.

\bibitem{Correa2011a}
C.~D. Correa and P.~Lindstrom.
\newblock Towards robust topology of sparsely sampled data.
\newblock {\em Transactions on Computer Graphics and Visualization},
  17(12):1852--1861, 2011.

\bibitem{DeBerg1997}
M.~de~Berg, M.~van Kreveld, M.~Overmars, and O.~Schwarzkopf.
\newblock {\em Computational Geometry, Algorithms and Applications}.
\newblock Springer, 3rd edition, 2008.

\bibitem{Desbrun2005a}
M.~Desbrun, E.~Kanso, and Y.~Tong.
\newblock Discrete differential forms for computational modeling.
\newblock In {\em {ACM SIGGRAPH} 2006 Courses}, pages 39--54. ACM, 2006.

\bibitem{Desbrun2006}
M.~Desbrun, K.~Polthier, P.~Schr\"oder, and A.~Stern.
\newblock Discrete differential geometry.
\newblock In {\em ACM SIGGRAPH 2006 Courses}, page~1. ACM, 2006.

\bibitem{Du1999}
Q.~Du, V.~Faber, and M.~Gunzburger.
\newblock Centroidal {V}oronoi tessellations: Applications and algorithms.
\newblock {\em SIAM Review}, 41(4):637--676, 1999.

\bibitem{Edelsbrunner2006}
H.~Edelsbrunner.
\newblock {\em Geometry and Topology for Mesh Generation}.
\newblock Cambridge Monographs on Applied and Computational Mathematics.
  Cambridge University Press, 2001.

\bibitem{EdelsbrunnerHarer2002}
H.~Edelsbrunner and J.~Harer.
\newblock Jacobi sets of multiple {M}orse functions.
\newblock In F.~Cucker, R.~DeVore, P.~Olver, and E.~S\"{u}li, editors, {\em
  Foundations of Computational Mathematics, Minneapolis 2002}, pages 37--57.
  Cambridge University Press, 2002.

\bibitem{EdelsbrunnerHarer2010}
H.~Edelsbrunner and J.~Harer.
\newblock {\em Computational Topology: An Introduction}.
\newblock American Mathematical Society, 2010.

\bibitem{EdelsbrunnerHarerNatarajan2003}
H.~Edelsbrunner, J.~Harer, V.~Natarajan, and V.~Pascucci.
\newblock {M}orse-{S}male complexes for piecewise linear 3-manifolds.
\newblock {\em Proceedings of the 19th {ACM} Symposium on Computational
  Geometry}, pages 361--370, 2003.

\bibitem{EdelsbrunnerHarerNatarajan2004}
H.~{Edelsbrunner}, J.~{Harer}, V.~{Natarajan}, and V.~{Pascucci}.
\newblock Local and global comparison of continuous functions.
\newblock In {\em IEEE Visualization}, pages 275--280, Oct 2004.

\bibitem{EdelsbrunnerHarerPatel2008}
H.~Edelsbrunner, J.~Harer, and A.~K. Patel.
\newblock Reeb spaces of piecewise linear mappings.
\newblock In {\em Proceedings of the 24th Annual Symposium on Computational
  Geometry}, pages 242--250. ACM, 2008.

\bibitem{EdelsbrunnerHarerZomorodian2003}
H.~Edelsbrunner, J.~Harer, and A.~J. Zomorodian.
\newblock Hierarchical {M}orse-{S}male complexes for piecewise linear
  2-manifolds.
\newblock {\em Discrete and Computational Geometry}, 30:87--107, 2003.

\bibitem{EdelsbrunnerLetscherZomorodian2002}
H.~Edelsbrunner, D.~Letscher, and A.~J. Zomorodian.
\newblock Topological persistence and simplification.
\newblock {\em Discrete \& Computational Geometry}, 28:511--533, 2002.

\bibitem{EdelsbrunnerMorozov2012}
H.~Edelsbrunner and D.~Morozov.
\newblock Persistent homology: Theory and practice.
\newblock {\em European Congress of Mathematics}, 2012.

\bibitem{EdelsbrunnerMorozov2017}
H.~Edelsbrunner and D.~Morozov.
\newblock Persistent homology.
\newblock In J.~E. Goodman, J.~O'Rourke, and C.~D. T\'{o}th, editors, {\em
  Handbook of Discrete and Computational Geometry}, Discrete Mathematics and
  Its Applications, chapter~24. CRC Press LLC, 2017.

\bibitem{Ester1996}
M.~Ester, H.-P. Kriegel, J.~Sander, and X.~Xu.
\newblock A density-based algorithm for discovering clusters a density-based
  algorithm for discovering clusters in large spatial databases with noise.
\newblock In {\em Proceedings of the 2nd International Conference on Knowledge
  Discovery and Data Mining}, pages 226--231. AAAI Press, 1996.

\bibitem{Estivill2002}
V.~Estivill-Castro.
\newblock Why so many clustering algorithms: A position paper.
\newblock {\em {ACM SIGKDD} Explorations Newsletter}, 4(1):65--75, June 2002.

\bibitem{fairman1997cie}
H.~S. Fairman, M.~H. Brill, H.~Hemmendinger, et~al.
\newblock How the {CIE} 1931 color-matching functions were derived from
  wright-guild data.
\newblock {\em Color Research \& Application}, 22(1):11--23, 1997.

\bibitem{Farin2002}
G.~Farin.
\newblock {\em Curves and Surfaces for CAGD: A Practical Guide}.
\newblock The Morgan Kaufmann Series in Computer Graphics. Morgan Kaufmann
  Publishers, 5th edition, 2002.

\bibitem{folland1995introduction}
G.~B. Folland.
\newblock {\em Introduction to Partial Differential Equations}.
\newblock Princeton University Press, 2nd edition, 1995.

\bibitem{Forman2001}
R.~Forman.
\newblock A user's guide to discrete {M}orse theory.
\newblock {\em S{\'e}minaire Lotharingien de Combinatoire}, 48, 2002.

\bibitem{Gabriel1969}
K.~R. Gabriel and R.~R. Sokal.
\newblock A new statistical approach to geographic variation analysis.
\newblock {\em Systematic Biology}, 18(3):259--278, 1969.

\bibitem{Georgii2007}
H.-O. Georgii.
\newblock {\em Stochastics: Introduction to Probability and Statistics}.
\newblock De Gruyter, 2008.

\bibitem{Ghrist2008}
R.~Ghrist.
\newblock Three examples of applied and computational homology.
\newblock {\em Nieuw Archief voor Wiskunde (The Amsterdam Archive, Special
  issue on the occasion of the fifth European Congress of Mathematics )}, pages
  122--125, 2008.

\bibitem{Glassner1995}
A.~S. Glassner.
\newblock {\em Principles of Digital Image Synthesis}.
\newblock The Morgan Kaufmann Series in Computer Graphics and Geometric
  Modeling. Morgan Kaufmann Publishers Inc., 1995.

\bibitem{Goodman2009}
R.~Goodman and N.~R. Wallach.
\newblock {\em Symmetry, Representations, and Invariants}.
\newblock Number 255 in Graduate Texts in Mathematics. Springer, 2009.

\bibitem{Gosink2007}
L.~{Gosink}, J.~{Anderson}, W.~{Bethel}, and K.~{Joy}.
\newblock Variable interactions in query-driven visualization.
\newblock {\em IEEE Transactions on Visualization and Computer Graphics},
  13(6):1400--1407, Nov 2007.

\bibitem{grassmann1853theorie}
H.~Grassmann.
\newblock Zur {T}heorie der {F}arbenmischung.
\newblock {\em Annalen der Physik}, 165(5):69--84, 1853.

\bibitem{guild1932colorimetric}
J.~Guild.
\newblock The colorimetric properties of the spectrum.
\newblock {\em Philosophical Transactions of the Royal Society of London.
  Series A}, 230:149--187, 1932.

\bibitem{Hansen2014}
C.~D. Hansen, M.~Chen, C.~R. Johnson, A.~E. Kaufman, and H.~Hagen, editors.
\newblock {\em Scientific Visualization: Uncertainty, Multifield, Biomedical,
  and Scalable Visualization}.
\newblock Mathematics and Visualization. Springer, 2014.

\bibitem{Hatcher2002}
A.~Hatcher.
\newblock {\em Algebraic Topology}.
\newblock Cambridge University Press, 2002.

\bibitem{huertas2006performance}
R.~Huertas, M.~Melgosa, and C.~Oleari.
\newblock Performance of a color-difference formula based on {OSA-UCS} space
  using small-medium color differences.
\newblock {\em JOSA A}, 23(9):2077--2084, 2006.

\bibitem{international2004colorimetry}
{International Commission on Illumination}.
\newblock {\em Colorimetry}.
\newblock CIE technical report. Commission Internationale de l'Eclairage, 2004.

\bibitem{Ip2012}
C.~Y. {Ip}, A.~{Varshney}, and J.~{JaJa}.
\newblock Hierarchical exploration of volumes using multilevel segmentation of
  the intensity-gradient histograms.
\newblock {\em IEEE Transactions on Visualization and Computer Graphics},
  18(12):2355--2363, Dec 2012.

\bibitem{Jain1999}
A.~K. Jain, M.~N. Murty, and P.~J. Flynn.
\newblock Data clustering: A review.
\newblock {\em ACM Computing Surveys}, 31(3):264--323, Sept. 1999.

\bibitem{James1999}
I.~M. James, editor.
\newblock {\em History of Topology}.
\newblock Elsevier B.V., 1999.

\bibitem{Jankowai2018}
J.~{Jankowai} and I.~{Hotz}.
\newblock Feature level-sets: Generalizing iso-surfaces to multi-variate data.
\newblock {\em IEEE Transactions on Visualization and Computer Graphics}, pages
  1--1, 2018.

\bibitem{judd1968ideal}
D.~B. Judd.
\newblock Ideal color space: Curvature of color space and its implications for
  industrial color tolerances.
\newblock {\em Palette}, 29(21-28):4--25, 1968.

\bibitem{judd1970ideal}
D.~B. Judd.
\newblock Ideal color space.
\newblock {\em Color Engineering}, 8(2):37, 1970.

\bibitem{Jungnickel2012}
D.~Jungnickel.
\newblock {\em Graphs, Networks and Algorithms}.
\newblock Algorithms and Computation in Mathematics. Springer, 4th edition,
  2012.

\bibitem{Kasten2016}
J.~Kasten, J.~Reininghaus, I.~Hotz, H.-C. Hege, B.~R. Noack, G.~Daviller, and
  M.~Morzynski.
\newblock Acceleration feature points of unsteady shear flows.
\newblock {\em Archives of Mechanics}, 68(1):55--80, 2016.

\bibitem{Kindlmann2014}
G.~Kindlmann and C.~Scheidegger.
\newblock An algebraic process for visualization design.
\newblock {\em IEEE Transactions on Visualization and Computer Graphics},
  20(12), 2014.

\bibitem{Kratz2013}
A.~Kratz, C.~Auer, M.~Stommel, and I.~Hotz.
\newblock Visualization and analysis of second-order tensors: Moving beyond the
  symmetric positive-definite case.
\newblock {\em Computer Graphics Forum - State of the Art Reports},
  32(1):49--74, 2013.

\bibitem{Kratz2011}
A.~Kratz, B.~Meyer, and I.~Hotz.
\newblock {A Visual Approach to Analysis of Stress Tensor Fields}.
\newblock In H.~Hagen, editor, {\em Scientific Visualization: Interactions,
  Features, Metaphors}, volume~2 of {\em Dagstuhl Follow-Ups}, pages 188--211.
  Schloss Dagstuhl--Leibniz-Zentrum fuer Informatik, Dagstuhl, Germany, 2011.

\bibitem{kuehnel2015differential}
W.~K{\"u}hnel.
\newblock {\em Differential Geometry: Curves - Surfaces - Manifolds}.
\newblock Student Mathematical Library. American Mathematical Society, 2015.

\bibitem{Lamb2014}
E.~Lamb.
\newblock What we talk about when we talk about holes.
\newblock Scientific American Blog Network, December 2014.

\bibitem{LiuMaljovecWang2017}
S.~Liu, D.~Maljovec, B.~Wang, P.-T. Bremer, and V.~Pascucci.
\newblock Visualizing high-dimensional data: Advances in the past decade.
\newblock {\em {IEEE} Transactions on Visualization and Computer Graphics},
  23(3):1249--1268, 2017.

\bibitem{luo2001development}
M.~R. Luo, G.~Cui, and B.~Rigg.
\newblock The development of the cie 2000 colour-difference formula: Ciede2000.
\newblock {\em Color Research \& Application}, 26(5):340--350, 2001.

\bibitem{mahy1994evaluation}
M.~Mahy, L.~Eycken, and A.~Oosterlinck.
\newblock Evaluation of uniform color spaces developed after the adoption of
  {CIELAB and CIELUV}.
\newblock {\em Color Research \& Application}, 19(2):105--121, 1994.

\bibitem{morton2005numerical}
K.~W. Morton and D.~F. Mayers.
\newblock {\em Numerical Solution of Partial Differential Equations: An
  Introduction}.
\newblock Cambridge University Press, 2005.

\bibitem{Munkres1984}
J.~R. Munkres.
\newblock {\em Elements of algebraic topology}.
\newblock CRC Press Taylor \& Francis Group, 1984.

\bibitem{Munzner2014}
T.~Munzner.
\newblock {\em Visualization Analysis \& Design}.
\newblock CRC Press Taylor \& Francis Group, 2014.

\bibitem{NagarajNatarajan2011}
S.~Nagaraj and V.~Natarajan.
\newblock Simplification of {J}acobi sets.
\newblock In V.~Pascucci, X.~Tricoche, H.~Hagen, and J.~Tierny, editors, {\em
  Topological Data Analysis and Visualization: Theory, Algorithms and
  Applications}, Mathematics and Visualization, pages 91--102. Springe, 2011.

\bibitem{OConnorRobertson1996}
J.~J. O'Connor and E.~F. Robertson.
\newblock {\em A History of Topology}.
\newblock MacTutor History of Mathematics, 1996.

\bibitem{Peacock2015}
T.~Peacock, G.~Froyland, and G.~Haller.
\newblock Introduction to focus issue: Objective detection of coherent
  structures.
\newblock {\em Chaos}, 25, 2015.

\bibitem{Peikert1999}
R.~Peikert and M.~Roth.
\newblock The ``parallel vectors" operator - a vector field visualization
  primitive.
\newblock {\em Proceedings of IEEE Visualization}, 14(16):263--270, 1999.

\bibitem{Polthier1998}
K.~Polthier and M.~Schmies.
\newblock Straightest geodesics on polyhedral surfaces.
\newblock In H.-C. Hege and K.~Polthier, editors, {\em Mathematical
  Visualization}, page 391. Springer Verlag, 1998.

\bibitem{Post1995}
F.~H. Post, F.~J. Post, T.~V. Walsum, and D.~Silver.
\newblock Iconic techniques for feature visualization.
\newblock In {\em Proceedings of the 6th Conference on Visualization}, page
  288. IEEE Computer Society, 1995.

\bibitem{Potter2010}
K.~Potter.
\newblock {\em The Visualization of Uncertainty}.
\newblock PhD thesis, University of Utah, 2010.

\bibitem{Press1992}
W.~Press, S.~Teukolsky, W.~Vetterling, and B.~Flannery.
\newblock {\em Numerical Recipes in {C}: The Art of Scientific Computing}.
\newblock Cambridge University Press, 3rd edition, 1992.

\bibitem{Raj2018}
M.~Raj.
\newblock {\em Depth-based visualizaitons for ensemble data and graphs}.
\newblock PhD thesis, University of Utah, 2018.

\bibitem{raj2012riemannian}
D.~Raj~Pant and I.~Farup.
\newblock Riemannian formulation and comparison of color difference formulas.
\newblock {\em Color Research \& Application}, 37(6):429--440, 2012.

\bibitem{Reeb1946}
G.~Reeb.
\newblock Sur les points singuliers d'une forme de pfaff completement
  intergrable ou d'une fonction numerique.
\newblock {\em Comptes Rendus Acad.Science Paris}, 222:847--849, 1946.

\bibitem{renardy2006introduction}
M.~Renardy and R.~C. Rogers.
\newblock {\em An introduction to Partial Differential Equations}, volume~13.
\newblock Springer Science \& Business Media, 2006.

\bibitem{resnikoff1974differential}
H.~L. Resnikoff.
\newblock Differential geometry and color perception.
\newblock {\em Journal of Mathematical Biology}, 1(2):97--131, 1974.

\bibitem{schrodinger1920grundlinien}
E.~Schr{\"o}dinger.
\newblock Grundlinien einer {T}heorie der {F}arbenmetrik im {T}agessehen.
\newblock {\em Annalen der Physik}, 368(22):481--520, 1920.

\bibitem{Schubert2017}
E.~Schubert, J.~Sander, M.~Ester, H.~P. Kriegel, and X.~Xu.
\newblock {DBSCAN} revisited, revisited: Why and how you should (still) use
  {DBSCAN}.
\newblock {\em ACM Transactions on Database Systems}, 42(3):19:1--19:21, July
  2017.

\bibitem{Schwichtenberg2017}
J.~Schwichtenberg.
\newblock {\em Physics from Symmetry}.
\newblock Undergraduate Lecture Notes in Physics. Springer, 2nd edition, 2017.

\bibitem{Shi2000}
J.~Shi and J.~{Malik}.
\newblock Normalized cuts and image segmentation.
\newblock {\em IEEE Transactions on Pattern Analysis and Machine Intelligence},
  22(8):888--905, Aug 2000.

\bibitem{Shirley2005}
P.~Shirley.
\newblock {\em Fundamentals of Computer Graphics}.
\newblock AK Peters, Ltd., 2005.

\bibitem{snider1987introduction}
A.~D. Snider and H.~F. Davis.
\newblock {\em Introduction to Vector Analysis}.
\newblock William C. Brown, 7th edition, 1987.

\bibitem{Sonka2008}
M.~Sonka, V.~Hlavac, and R.~Bohle.
\newblock {\em Image processing, analysis and and machine vision}.
\newblock Thomson, 3rd edition, 2008.

\bibitem{Steward2019}
J.~Steward.
\newblock {\em Multivariate Calculus}.
\newblock Brooks/Cole CENGAGE Learning, 7th edition, 2019.

\bibitem{Strang2016}
G.~Strang.
\newblock {\em Introduction to Linear Algebra}.
\newblock Wellesley-Cambridge Press, 5th edition, 2016.

\bibitem{Suthar2013}
N.~Suthar, I.~jeet Rajput, and V.~kumar Gupta.
\newblock A technical survey on {DBSCAN} clustering algorithm.
\newblock {\em International Journal of Scientific and Engineering Research},
  4(5), 2013.

\bibitem{Telea2015}
A.~C. Telea.
\newblock {\em Data Visualization: Principles and Practice}.
\newblock AK Peters, Ltd., 2nd edition, 2015.

\bibitem{Tufte2001}
E.~R. Tufte.
\newblock {\em The Visual Display of Quantitative Information}.
\newblock Graphics Press, 2001.

\bibitem{urban2007embedding}
P.~Urban, M.~R. Rosen, R.~S. Berns, and D.~Schleicher.
\newblock Embedding non-{Euclidean} color spaces into euclidean color spaces
  with minimal isometric disagreement.
\newblock {\em Journal of the Optical Society of America A}, 24(6):1516--1528,
  2007.

\bibitem{von1867handbuch}
H.~Von~Helmholtz.
\newblock {\em Handbuch der physiologischen Optik}, volume~9.
\newblock Voss, 1867.

\bibitem{WangJ2018}
J.~Wang, S.~Hazarika, C.~Li, and H.-W. Shen.
\newblock Visualization and visual analysis of ensemble data: A survey.
\newblock {\em {IEEE} Transactions on Visualization and Computer Graphics},
  2018.

\bibitem{Wardetzky2007}
M.~Wardetzky, S.~Mathur, F.~K{\"a}lberer, and E.~Grinspun.
\newblock {D}iscrete {L}aplace operators: {N}o free lunch.
\newblock In {\em Proc. Eurographics Symposium on Geometry processing}, pages
  33--37, 2007.

\bibitem{Weyl1952}
H.~Weyl.
\newblock {\em Symmetry}.
\newblock Princeton University Press, 1952.

\bibitem{Whitaker2013}
R.~T. Whitaker, M.~Mirzargar, and R.~M. Kirby.
\newblock Contour boxplots: A method for characterizing uncertainty in feature
  sets from simulation ensembles.
\newblock {\em IEEE Transactions on Visualization and Computer Graphics},
  19(12):2713--2722, Dec. 2013.

\bibitem{WikipediaTopology2018}
{Wikipedia contributors}.
\newblock Topology.
\newblock Wikipedia, The Free Encyclopedia, March 2018.

\bibitem{Wong2000}
P.~C. {Wong}, H.~{Foote}, R.~{Leung}, D.~{Adams}, and J.~{Thomas}.
\newblock Data signatures and visualization of scientific data sets.
\newblock {\em IEEE Computer Graphics and Applications}, 20(2):12--15, March
  2000.

\bibitem{Woodring2009}
J.~{Woodring} and H.~{Shen}.
\newblock Multiscale time activity data exploration via temporal clustering
  visualization spreadsheet.
\newblock {\em IEEE Transactions on Visualization and Computer Graphics},
  15(1):123--137, Jan 2009.

\bibitem{Wu2013}
M.~Q.~Y. {Wu}, R.~{Faris}, and K.~{Ma}.
\newblock Visual exploration of academic career paths.
\newblock In {\em {IEEE/ACM} International Conference on Advances in Social
  Networks Analysis and Mining}, pages 779--786, Aug 2013.

\bibitem{wyszecki1982color}
G.~Wyszecki and W.~S. Stiles.
\newblock {\em Color Science}, volume~8.
\newblock Wiley New York, 1982.

\bibitem{zeyen2018interpolation}
M.~Zeyen, T.~Post, H.~Hagen, J.~Ahrens, D.~Rogers, and R.~Bujack.
\newblock Color interpolation for non-{Euclidean} color spaces.
\newblock In {\em IEEE Scientific Visualization Conference Short Papers}. IEEE,
  2018.

\bibitem{ZhangYJ2016}
Y.~J. Zhang.
\newblock {\em Geometric Modelng and Mesh Generation from Scanned Images}.
\newblock CRC Press Taylor \& Francis Group, 2016.

\end{thebibliography}

\end{document}